# High Abundances of Presolar Grains and $^{15}N$-rich Organic Matter in CO3.0 Chondrite Dominion Range 08006

Larry R. Nittler[1], Conel M. O'D. Alexander[1], Jemma Davidson[1], My E. I. Riebe[1], Rhonda M. Stroud[2], and Jianhua Wang[1]

[1]Department of Terrestrial Magnetism, Carnegie Institution of Washington, Washington, DC 20015, USA.

[2]Materials Science and Technology Division, Code 6366, US Naval Research Laboratory, Washington, DC 20375-5320, USA.

## Abstract

NanoSIMS C-, N-, and O-isotopic mapping of matrix in CO3.0 chondrite Dominion Range (DOM) 08006 revealed it to have in its matrix the highest abundance of presolar O-rich grains (257 +76/-96 ppm, 2$\sigma$) of any meteorite. It also has a matrix abundance of presolar SiC of 35 (+25/-17, 2$\sigma$) ppm, similar to that seen across primitive chondrite classes. This provides additional support to bulk isotopic and petrologic evidence that DOM 08006 is the most primitive known CO meteorite. Transmission electron microscopy of five presolar silicate grains revealed one to have a composite mineralogy similar to larger amoeboid olivine aggregates and consistent with equilibrium condensation, two non-stoichiometric amorphous grains and two olivine grains, though one is identified as such solely based on its composition. We also found insoluble organic matter (IOM) to be present primarily as sub-micron inclusions with ranges of C- and N-isotopic anomalies similar to those seen in primitive CR chondrites and interplanetary dust particles. In contrast to other primitive extraterrestrial materials, H isotopic imaging showed normal and homogeneous D/H. Most likely, DOM 08006 and other CO chondrites accreted a similar complement of primitive and isotopically anomalous organic matter to that found in other chondrite classes and IDPs, but the very limited amount of thermal metamorphism experienced by DOM 08006 has caused loss of D-rich organic moieties, while not substantially affecting either the molecular carriers of C and N anomalies or most inorganic phases in the meteorite. One C-rich grain that was highly depleted in $^{13}C$ and $^{15}N$ was identified; we propose it originated in the Sun's parental molecular cloud.

## 1. INTRODUCTION

Chondritic meteorites and micrometeorites, stratospheric interplanetary dust particles (IDPs) and dust from comet Wild 2 brought to Earth by NASA's Stardust spacecraft all contain traces of materials that pre-date the formation of the Sun's protoplanetary disk. Presolar stardust grains originated as condensates in the outflows and ejecta of evolved stars. They are recognized by their highly anomalous isotopic compositions, which largely reflect nuclear processes that occurred in the grains' parent stars (Zinner, 2014; Nittler and Ciesla, 2016). Additionally, a large fraction of the C present in primitive meteoritic materials is in the form of a macromolecular organic matter ("insoluble organic matter" or IOM) that shows isotopic anomalies (typically high D/H and/or $^{15}N/^{14}N$ ratios relative to terrestrial values) strongly indicative of an origin in either the Sun's parental molecular cloud, or in the outer reaches of the nascent Solar System, where conditions were similar to those in molecular clouds (Messenger, 2000; Busemann et al., 2006).

A large number of presolar stardust phases have been identified, the most abundant being crystalline and amorphous silicates, $Al_2O_3$, $MgAl_2O_4$, and SiC, but also including graphitic grains (often containing sub-grains of metal and carbides), $Si_3N_4$, and other minor phases (Zinner, 2014). Nanometer-sized diamonds (nanodiamonds) carrying isotopically anomalous noble gases are also abundant, but their small sizes make their identification as presolar grains ambiguous (Dai et al., 2002; Stroud et al., 2011; Heck et al., 2014). Presolar grains are essentially fossil remnants of stars that survived interstellar and solar nebular processing, and as such they have been used as tools to probe a wide variety of astrophysical processes. Because they encompass such a wide range of chemical forms, they also can serve as sensitive probes of processes that affected materials in the interstellar medium (e.g., shocks and irradiation, Vollmer et al., 2007), the protoplanetary disk (e.g., heating, Huss and Lewis, 1995), and in asteroidal and cometary parent bodies (e.g., aqueous alteration, thermal metamorphism: Floss and Stadermann, 2012; Leitner et al., 2012b). In particular, detailed analyses of presolar grain abundances in meteorites suggest that the different chondrite groups accreted a similar mix of presolar materials and abundance variations observed today reflect that different phases respond differently to parent-body processes. For example, the low-temperature aqueous alteration that has affected many carbonaceous chondrites has little or no obvious impact on SiC abundances (Davidson et al., 2014a), but can be very destructive of presolar silicates (Leitner et al., 2012b). Thermal metamorphism destroys all presolar grains, but at different rates. For instance, presolar $MgAl_2O_4$ is in much higher abundance in meteorites that have seen lower degrees of heating, like CM chondrites (Zinner et al., 2003) and the LL3 Krymka (Nittler et al., 2008), than in the more heated Tieschitz (H/L3.6) ordinary chondrite (Nittler et al., 1997). Of the known stardust phases, silicates are the most susceptible to destruction by parent-body processing and their abundances (and the ratio of their abundances to those of the more resistant oxide phases) are thus highly sensitive indicators of the degree to which the material they are embedded in has been processed.

The highest presolar silicate abundances have been observed in anhydrous fine-grained IDPs collected in Earth's stratosphere and micrometeorites collected in Antarctica, with reported abundances ranging from a few hundred ppm to percent levels (e.g., Floss et al., 2006; Busemann et al., 2009; Davidson et al., 2012; Alexander et al., 2017c). Floss et al. (2006) calculated an average presolar silicate abundance of 375 ppm for a sub-set of $^{15}N$-rich IDPs, termed "isotopically primitive." Anhydrous IDPs are suspected to have originated from comets and the high abundance of presolar grains supports this hypothesis, as comets are expected to be more primitive than the asteroidal parent bodies of chondritic meteorites. Prior to the present study, the highest presolar silicate abundances in meteorites have been reported for the highly primitive ungrouped carbonaceous chondrite Acfer 094 (150-200 ppm; Nguyen et al., 2007; Vollmer et al., 2009b; Hoppe et al., 2015), the CO 3.0 chondrite Allan Hills (ALH) 77307 (190 ppm; Nguyen et al., 2010), and the relatively unaltered CR2 chondrites Queen Alexandra Range (QUE) 99177 and Meteorite Hills (MET) 00426 (160-220 ppm; Floss and Stadermann, 2009a; Nguyen et al., 2010). Because presolar grains reside in the fine-grained matrix between chondrules and other inclusions in chondrites, these quoted abundances are all matrix-normalized, but are still lower than the average abundance reported for IDPs. These lower abundances indicate that there has been significant destruction of presolar silicates either in the nebular regions where the chondrites accreted or during the very minor degrees of alteration that petrographic studies indicate occurred in the parent bodies of these four meteorites (Brearley, 1993; Greshake, 1997; Harju et al., 2014).

An additional type of primitive, possibly presolar, material found in meteorites and IDPs is isotopically anomalous organic matter (Messenger, 2000; Busemann et al., 2006; Alexander et al., 2007; Davidson et al., 2012; Alexander et al., 2017). Macromolecular insoluble organic matter (IOM) is the dominant form of C in primitive chondrites and IDPs. Large isotopic variations, most notably enrichments in D and $^{15}N$, relative to terrestrial D/H and $^{15}N/^{14}N$ ratios, are commonly observed in the bulk isotope compositions for organic matter from different parent bodies, and even greater variations often occur at the μm to sub-μm scale in the most primitive samples (Messenger, 2000; Keller et al., 2004; Busemann et al., 2006; Alexander et al., 2007; Davidson et al., 2012; De Gregorio et al., 2013). Carbon-isotopic anomalies are also seen occasionally in IOM from IDPs and primitive meteorites (Floss et al., 2004; Busemann et al., 2006; Floss and Stadermann, 2009b). *In situ* studies (e.g., Busemann et al., 2006; Remusat et al., 2010; Bose et al., 2014; Le Guillou and Brearley, 2014) of chondrites indicate that IOM is generally present as discrete, typically sub-μm, grains, including "nanoglobules," which are spherical and often hollow solid organic grains (Nakamura-Messenger et al., 2006; De Gregorio et al., 2013). How the meteoritic IOM formed is still an open question. The large D and $^{15}N$ enrichments as well as infrared spectral similarities with dust in the diffuse interstellar medium have long been interpreted to indicate an interstellar heritage for the IOM (Robert and Epstein, 1982; Yang and Epstein, 1983; Pendleton et al., 1994), or that it formed in parent bodies from originally interstellar material (Cody et al., 2011; Vollmer et al., 2014). Alternatively, it has been suggested that the IOM may have formed in the solar nebula, e.g., by irradiation of simpler molecular ice precursors or by Fischer-Tropsch type processes (Gourier et al., 2008; Nuth et al., 2008; Ciesla and Sandford, 2012). Complementary to presolar grain abundances, the nature of IOM, e.g. abundance, isotopic composition, and chemical structure, can also be a sensitive indicator of thermal metamorphism and aqueous alteration in meteorite parent asteroids (Quirico et al., 2003; Alexander et al., 2007; Busemann et al., 2007; Cody et al., 2008; Alexander et al., 2013; Bonal et al., 2016; Alexander et al., 2018).

In this paper, we report the discovery that the Antarctic CO3 chondrite Dominion Range (DOM) 08006 has a higher matrix-normalized abundance of O-rich presolar stardust grains than any previously studied chondrite. Moreover, we found DOM 08006 to contain abundant organic material exhibiting a similar range of C- and N-isotopic anomalies to that seen in primitive CR and CM chondrites (albeit with much more muted D anomalies than seen in those meteorite classes). That DOM 08006 may be special was indicated by a prior study of amino acids in a number of CO chondrites by Burton et al. (2012). These authors found that it contained a lower abundance of straight-chain amino acids (associated with thermal alteration) than ALH 77307, suggesting that DOM 08006 may have experienced less heating than ALH 77307, hitherto considered the least altered of the CO3 chondrites (Grossman and Brearley, 2005). Spurred on by this work, Alexander et al. (2018) included DOM 08006 in a survey of bulk isotopic and mineralogical properties of primitive CO chondrites and found that this meteorite has a significantly higher C content and slightly higher H/C, D, and $^{15}N$ contents compared to ALH 77307, indicating that it has preserved more primitive organic matter. The high abundance of presolar materials reported here, together with detailed bulk chemical and isotopic, mineralogical and petrologic data for this meteorite reported by Davidson et al. (2014b) and Alexander et al. (2018), bears this out and indicates that DOM 08006 is among the least altered of any chondritic meteorite. It is thus a very valuable specimen both for understanding the original unaltered nature of nebular materials that accreted onto planetesimals in the protoplanetary disk as well as the earliest stages of alteration on asteroids.

## 2. SAMPLE AND EXPERIMENTAL METHODS

We obtained a polished thin section (DOM 08006, 16) from the Johnson Space Center (Figure 1) and analyzed it by isotopic imaging with the Cameca NanoSIMS 50L ion microprobe at the Carnegie Institution of Washington. We used very similar methods to those used previously for presolar grain and organic mapping of meteorite thin sections (Nguyen et al., 2010). For most measurements, we rastered a focused $Cs^+$ primary ion beam over 10μm×10μm areas with simultaneous collection of negative secondary ions of the C isotopes (either $^{12,13}C$ or $^{12}C_2$, $^{12}C^{13}C$), $^{16,17,18}O$, $^{28}Si$, either $^{12}C^{14}N$ or $^{27}Al^{16}O$ on individual electron multipliers (EMs) along with secondary electrons. The effective spatial resolution (taking into account beam size and slight broadening due to inaccuracy in correcting for stage/beam shifts over the long measurements) for most images was between 100 and 150 nm. One set of images was acquired in $^{16,17}O$, $^{12}C_2$, $^{12}C^{13}C$, $^{12}C^{14}N$, $^{12}C^{15}N$, $^{28}Si$ negative ions and secondary electrons to characterize the N isotopic composition of carbonaceous material. For all measurements, the mass resolving power was set so that isobaric interferences (e.g., $^{16}OH$ on $^{17}O$, $^{13}C_2$ on $^{12}C^{14}N$) were negligible and the primary beam current was set (~1.2 pA) such that the maximum $^{16}O$ count rate in each image was typically

~$10^6$ counts per second. Twenty sequential 256×256 pixel images (cycles) were acquired on each area, for a total acquisition time of 65 minutes (total dwell time of 60 milliseconds per 39 nm × 39 nm pixel). A Nuclear Magnetic Resonance probe was used to maintain magnetic field stability in the mass spectrometer, and the peak positions of $^{12}$C and $^{16}$O were checked every five cycles and the other peaks shifted accordingly. Prior to image acquisition, areas were pre-sputtered for roughly three minutes with a more intense primary beam to remove the C coat and implant Cs, which greatly boosts the yield of negative secondary ions.

In addition, two sets of imaging runs on the DOM 08006 thin section were carried out with a higher (~10 pA) $Cs^+$ beam current to analyze H isotopes together with those of C ($^{1,2}$H, $^{12,13}$C, $^{16}$O, and $^{12}$C$^{14}$N). The spatial resolution was degraded to ~350 nm due to the higher primary beam current, and slightly longer dwell times were also used (100 ms total counting time per pixel).

We used the L'IMAGE software (L. R. Nittler) to analyze the image data. Images were corrected for the 44 ns deadtime of the EM counting system and shifts between frames were corrected for via an autocorrelation algorithm. We also corrected the O-isotopic images for quasi-simultaneous arrival (QSA) effects (Slodzian et al., 2004). QSA refers to the possibility that a single primary ion may produce two or more secondary ions of a given species within a time interval much shorter than the deadtime of the NanoSIMS EM counting system. Since the probability of this occurring is much higher for $^{16}$O than for $^{17}$O or $^{18}$O, especially with the high secondary ion transmission of the NanoSIMS, this can lead to undercounting of $^{16}$O and hence introduce an isotopic fractionation. Slodzian et al. (2004) showed that, according to Poisson statistics, the magnitude of the fractionation should be proportional to the ratio of secondary ion to primary ion count rates, but empirically the proportionality constant does not in general match the prediction. We applied the formalism of Slodzian et al. (2004) to $^{18}$O/$^{16}$O ratio images to empirically determine and correct $^{16}$O images for the appropriate QSA proportionality constant. We generally calculated the constant from the individual pixels of the first image acquired in an area of the sample and used it to correct all the images for that run. The maximum correction for a given image pixel was typically less than 10%. Both the deadtime and QSA corrections were carried out on a pixel-by-pixel basis on individual image frames.

Following the deadtime, QSA and alignment corrections, we calculated pixel-by-pixel isotopic ratio images (e.g., $^{17}$O/$^{16}$O, $^{18}$O/$^{16}$O, $^{13}$C/$^{12}$C, $^{15}$N/$^{14}$N, D/H) after first applying a 3×3 pixel boxcar smoothing to the images to improve the signal to noise ratio and to remove spurious variations below the true spatial resolution. Oxygen isotopes were internally normalized to the average composition of each image, H and C isotopes were normalized to epoxy infilling cracks in the thin section. Comparison to measurements of IOM from the CR chondrite QUE 99177 used as an external standard indicated that the epoxy has $\delta$D and $\delta^{13}$C within 10% and 5% of zero, respectively. Nitrogen isotopes were normalized so that the average composition of carbonaceous particles in an image matched the bulk value of $\delta^{15}$N=10.5 ‰ for IOM from this meteorite (Alexander et al., 2018). Presolar O-rich and C-rich grain candidates were identified by manual examination of the isotopic ratio images and of "sigma" images, in which each pixel represents the number of standard deviations (based on Poisson statistics) its measured isotopic ratio is away from terrestrial values (Figure 2). Candidate grain regions of interest (ROIs) were selected as regions of several contiguous pixels clearly deviating from the standard ratio(s). In general, only pixels within the full-width at half maximum of the anomalous region in a "sigma" image were included in an ROI (Fig. 2). Isotopic ratios were then calculated from the summed counts of all pixels within an ROI, which was confirmed to be a presolar grain candidate if at least one isotopic ratio was >3$\sigma$ away from the surrounding meteorite matrix. Some candidates, including all grains with anomalous $^{17}$O/$^{16}$O identified during the O-N runs as well as several grains for which the apparent anomaly was marginal (e.g., <4$\sigma$), were subsequently re-measured for all three O isotopes, again in imaging mode but with a smaller raster, typically 3µm×3µm. To characterize organic materials, we used an automatic particle definition algorithm (Nittler et al., 1997) to define individual C-rich ROIs in the images and calculated C-, N- and/or H-isotopic ratios from the summed pixel counts. This was only done for the combined C and N and combined H and C runs. Note that because the NanoSIMS primary ion beam tails will always contribute some signal from surrounding material to the measurement of grains of interest, all reported isotopic anomalies are lower limits.

Following the NanoSIMS measurements, the imaged areas were examined in a JEOL 6500F scanning electron microscope (SEM) equipped with an energy dispersive X-ray analysis system.

Four cross-sections sampling five presolar silicate grains were extracted by focused ion beam lift-out with a FEI Nova 600 FIB-SEM at the Naval Research Laboratory (NRL) for analysis by transmission electron microscopy (TEM). Analytical TEM studies were performed with the JEOL 2200FS field-emission scanning transmission electron microscope at NRL, equipped with a Noran System Six energy dispersive X-ray spectrometer. EDS spectra of individual grains were quantified with Cliff-Lorimer routines, with K factors calibrated from San Carlos olivine and Tanzanian hibonite standards.

## 3. RESULTS

### 3.1. NanoSIMS imaging

We mapped four areas of DOM 08006 fine-grained matrix that were separated from one another by several mm (Fig. 1) in case of heterogeneity across the section. A total of ~28,000 μm$^2$ was covered for O and C isotopes, including about 700 μm$^2$ measured for N isotopes as well (Section 2). A total of 6200 μm$^2$ were covered for H isotopes. Some details of the mapping results are provided in Table 1. We note that following the acquisition of O and C data for Area 7, we found that the primary beam had inadvertently been slightly defocused so that the lateral resolution for these images was ~300 nm, compared to the resolution of 100-150 nm for the other runs. Example images of one particularly interesting 10 × 10 μm$^2$ region in Area 3 containing a presolar silicate grain, a presolar SiC grain and isotopically anomalous organic material, are shown in Figure 3.

#### *3.1.1. O-rich presolar grains*

Analysis of the NanoSIMS images revealed a total of 101 grains with substantial anomalies (>4 σ) in their O-isotopic ratios indicating that they are presolar, circumstellar grains. Data for these are summarized in Table 2. Grain diameters range from 150 nm up to 1 μm (Figure 4). For most grains, these were estimated from the NanoSIMS images. If $N$ pixels were included in a grain ROI, the equivalent diameter is calculated from $2 \times \sqrt{(N \times \frac{a}{\pi})}$, where $a$ is the area of a single image pixel, equal to 0.00152 μm$^2$ for most of the images. However, the apparent size of isotopically anomalous regions in NanoSIMS images depends both on the intrinsic size of the anomalous grain and the size of the approximately Gaussian-shaped primary ion beam. We thus report two sizes for the grains in Table 2: the original size inferred from the images and a corrected size, where an assumed 120-nm beam diameter is subtracted in quadrature from the original diameter. In almost all previous presolar grain studies of meteorites and IDPs, only the former values are reported. The one exception is the study of Zhao et al. (2013), who applied an analogous correction to grain sizes determined by NanoSIMS based on comparison with images from the Auger microprobe, although their correction was apparently slightly larger, 20-30%, than ours. The average measured grain diameter is 293 nm, whereas this is decreased in our study by ~10% to 264 nm for the corrected data. We consider the corrected grain sizes to be conservative lower limits.

With the exception of the five grains for which we have TEM data (Section 3.2), we have little mineralogical information about the identified presolar O-rich grains. We can glean some elemental information from the NanoSIMS data. For 65 of the presolar grains, the NanoSIMS imaging runs included measurement of both $^{28}Si^{-}$ and $^{27}Al^{16}O^{-}$ secondary ions. The measured $Si^{-}/O^{-}$ secondary ion ratios are plotted against the $AlO^{-}/O^{-}$ ion ratios for these grains in Figure 5b. All of the grains show Si/O ion ratios in the range we typically observe for silicates. However, since the vast majority of known presolar oxide grains are Al-rich (Nittler et al., 2008), we assumed that the grains with high AlO/O ratios are oxides and, based on a break in the histogram of AlO/O values (Figure 5a), chose a threshold of AlO/O=0.007, above which a grain may be identified as an oxide. Most of these grains have Si/O ratios at the lower end of the observed range, consistent with their measured Si being contributed from surrounding material during the NanoSIMS imaging measurements. Consequently, as an additional criteria for identifying oxide grains, we included only those grains with the lowest Si/O ratios, i.e., below 0.01 (grey box in Figure 5). With this threshold, 10 of the 65 grains are inferred to be oxides for an overall ratio of presolar silicates to oxides of 5.5. The uncertainty on this number is obviously quite large, however. For example, changing the Al/O threshold to 0.01 would raise the presolar silicate to presolar oxide ratio to ~12. Moreover, Nguyen et al. (2010) found that 6 of 10 presolar grains identified as oxides from NanoSIMS elemental images were subsequently revealed to be silicates by Auger microprobe analysis, and our inferred silicate/oxide ratio is thus likely a lower limit.

The O-isotopic ratios for the 101 identified O-rich presolar grains are compared with previous data for presolar oxides and silicates in Figures 6 and 7. It has long been recognized that the data for presolar silicates show narrower ranges of O isotopic ratios compared to the presolar oxide database (e.g., Nguyen et al., 2007). This is a direct result of the fact that the presolar silicates have almost exclusively been found by *in situ* isotopic imaging methods, whereas a large number of the oxides were measured as single isolated grains in meteoritic acid residues. Even for nominally resolved grains, the tails of the NanoSIMS primary ion beam contribute some atoms from isotopically normal surrounding material to the measurements of the presolar grains (Nguyen et al., 2007), a phenomenon usually referred to as "isotopic dilution." This has a larger effect on grains with $^{17}O$ and/or $^{18}O$ depletions, and the lack of presolar silicates with $^{18}O/^{16}O$ ratios less than about $5{\times}10^{-4}$, compared to the presolar oxides (Figures 6 and 7), is thus almost certainly an effect of isotope dilution, not a sign that such silicate grains are not present in the meteorites.

The DOM 08006 presolar grains are in generally good agreement with the previous data, but show slightly narrower ranges of O-isotope ratios compared to the larger presolar silicate database (bottom panels of Figure 7). It is unlikely that this represents a true difference in the isotopic compositions of presolar grains in this meteorite. More likely, it indicates both the more limited statistics of the present study and that isotopic dilution played a larger role in our O-isotopic measurements than in some previous studies. For all of the imaged areas except Area 7, our spatial resolution of 100–150 nm was comparable to that used in most other NanoSIMS in situ studies (but see Hoppe et al., 2015, 2017 for studies based on higher resolution). However, the measured isotopic composition of a grain analyzed in the NanoSIMS imaging mode depends critically on which pixels are included in the ROI calculation; including more pixels will generally increase statistical precision but also enhance the effect of isotope dilution. Few published studies of presolar silicates have specified what criteria were used to include pixels in the isotope ratio calculations. We included all pixels within the full width at half maximum of "sigma" images (Section 2, Fig. 2) and for many of the grains, the derived ratios would have been more anomalous had we chosen to restrict the calculation to fewer pixels. Therefore, it is likely that the narrower span of O isotopic ratios measured in DOM 08006 grains compared to the prior presolar silicate database is a result of our approach to defining grain ROIs. However, we note that essentially all but the largest grains are compromised by isotope dilution in all *in situ* studies (Nguyen et al., 2007; Nguyen et al., 2010), regardless of image processing choices, and almost all of the grains thus have true O-isotope anomalies that are more extreme than reported.

*3.1.2. C- and N-isotopic mapping*

We identified a total of 66 sub-μm carbonaceous inclusions with anomalous $^{13}C/^{12}C$ and/or $^{15}N/^{14}N$ ratios (Table 3). An ROI was identified as anomalous if its C isotopic ratio was more than 5% different from the terrestrial value (since almost all Solar System materials are within 5% of terrestrial) at a significance of at least 2σ or if its N isotopic ratio was more than 3σ away from the average value of DOM 08006 IOM ($\delta^{15}N$ =10.5 ‰; Alexander et al., 2018). Seventeen $^{13}C$-rich ROIs were identified as presolar SiC grains (e.g., Figure 3), based on the spatially correlated presence of Si. An additional four highly $^{13}C$-rich grains were identified during the H- and C-imaging runs (Section 3.1.3) and are probably SiC, but these measurements did not include Si and the poorer spatial resolution for these measurements makes their true sizes and isotopic compositions uncertain. The 17 SiC grains range in diameter from about 100 nm to 400 nm and their measured $^{12}C/^{13}C$ ratios range from 27 to 79. These $^{12}C/^{13}C$ ratios are compared to those of ~16,000 SiC grains from the literature in Figure 8. The literature data are divided into those measured as isolated single grains in meteorite acid residues (upper panel) and those measured in situ as in this study (lower panel). Although the statistics are limited, the peak of the DOM 08006 SiC histogram plots at $^{12}C/^{13}C$~70, compared to the peak of ~60 observed for the literature data for isolated SiC grains. A similar shift is seen for the literature in situ data, indicating that it is likely due to moderate isotopic dilution of the measured C isotopes by isotopically normal material (e.g., contamination, epoxy used to prepare the meteorite section, or intrinsic organic matter).

The rest of the C- and/or N-anomalous regions are not spatially associated with Si and are most likely particles of insoluble organic matter (IOM), though we cannot rule out that some are presolar graphite grains (Haenecour et al., 2016). The identified C- and/or N-anomalous grains show a similar range of grain sizes as the SiC grains and both enrichments and depletions are seen in $^{13}C$ and $^{15}N$ (Table 3). The $\delta^{13}C$ values of the C-anomalous grains of this study are compared to those seen in the highly primitive CR chondrites QUE 99177 and MET 00426 (Floss and Stadermann, 2009b) and CO3.0 ALH 77307 (Bose et al., 2012) in Figure 9. The data for all of the meteorites are consistent with two peaks in $\delta^{13}C$, with values of about ±150 ‰. However, we note that typically grains with $^{13}C/^{12}C$ ratios within ~10% of terrestrial are by definition not counted as anomalous in surveys such as ours. Thus, it is likely that the peaks in the histogram indicate that the meteoritic organic grains mostly have $\delta^{13}$C values between -200 and +200 ‰, but only the edges of the distribution are identified as anomalous.

The $\delta^{13}$C and $\delta^{15}$N values for the 36 C- and/or N-anomalous organic grains of this study for which both elements were measured are shown in Figure 10. Also shown are the "normal" grains from this study (those within 10% of terrestrial values within 3$\sigma$ for both isotopic ratios) along with data for similar anomalous grains measured in other primitive planetary materials. The C and N data are not well correlated, and the DOM 08006 grains span similar ranges to those previously seen in MET 00426 and ALH 77307. One difference is a lack of grains in DOM 08006 with $\delta^{15}$N values higher than 1000 ‰, in particular grains with $\delta^{15}$N≈1000–1500 ‰ and depleted $^{13}$C ($\delta^{13}$C~ -200 to -100 ‰), as seen in other meteorites as well as in some IDP and comet Wild 2 samples (dashed ellipse in Fig. 10).

One particularly interesting grain, A7_60, is highly depleted in both $^{13}$C and $^{15}$N (Table 3, Figures 10 and 11). Grain A7_60 appears as two adjacent C-rich regions in NanoSIMS and SEM images (Fig. 11), but the identical compositions of the two regions indicate that they were likely originally part of the same inclusion, that was either split during the polishing of the thin section or by the NanoSIMS beam, or is connected below the surface of the polished thin section. An attempted measurement of the H isotopes in this grain was unsuccessful due to very low secondary ion counts.

### *3.1.3. H-isotopic mapping*

A total of 2,400 μm$^2$ of Area 2 and 3,800 μm$^2$ of Area 7 were mapped for H and C isotopes as described in Section 2. Hydrogen was clearly correlated with C in all of the images as expected both for IOM and the epoxy used to make the thin section. However, on average the D/H ratio of the meteoritic C grains was indistinguishable from both that of the epoxy and of the terrestrial standard we used and we found no evidence for spatial heterogeneity in D/H (Figure 12). DOM 08006 has a bulk $\delta$D value of about 0 ‰, but a purified IOM residue from this meteorite shows a moderate bulk D enrichment ($\delta$D=476±25 ‰) and H/C=0.476 (Alexander et al., 2018). For our imaging measurements, we obtained typical errors of ~200–500 ‰ for individual ~500-nm sized carbonaceous grains and therefore we would not necessarily expect to resolve a 500 ‰ D enrichment for an individual grain. However, it is somewhat surprising not to see even a bulk enrichment when a large number of grains are averaged to obtain smaller statistical uncertainties. Moreover, our results are in strong contrast to D/H imaging results from other primitive carbonaceous chondrites, which have revealed highly heterogeneous D/H ratios, with δD values in individual sub-μm grains (hotspots) reaching into the tens of thousands (Busemann et al., 2006; Remusat et al., 2010). For example, in Figure 12 we compare NanoSIMS C and D/H images for same-sized areas of DOM 08006 (panels a and c) and the primitive CR chondrite QUE 99177 (panels b and d). The measurement conditions were nearly identical for these images, as can be seen from the similar $^{12}$C ion count rates (Figure 12a,c); H count rates (not shown) were likewise similar. However, whereas QUE 99177 shows D hotspots with δD values ranging from 5,000 to 15,000 ‰, the total range of D/H in the DOM 08006 image is small and none of the apparent isotopic variations are statistically significant. QUE 99177 IOM has a H/C=0.803 (Alexander et al., 2007), much higher than that measured for DOM 08006 IOM. Since the two samples showed similar H/C secondary ion ratios, we therefore conclude that the isotopically normal D/H we measure on average in DOM 08006, compared to the known D-enriched composition of purified IOM, is most likely due to H contamination in the sample. However, such contamination would not be expected to mask hotspots with much higher intrinsic D/H ratios.

To investigate the apparent lack of D hotspots in DOM 08006, we obtained D/H images of particles of IOM from both DOM 08006 and QUE 99177, pressed into gold foils as done in previous work (e.g., Busemann et al., 2006). The imaging results for DOM 08006 IOM (Figure 12e) reproduce the bulk measured δD of ~500 ‰ and, as in the *in situ* results, show no evidence for highly D-enriched hotspots, whereas these are abundant in the QUE 99177 data (Figure 12f). The combined *in situ* and IOM results provide further support that the *in situ* H-isotope data are affected by contamination and show that organic matter in DOM 08006 is much more isotopically uniform in D/H than is seen in many other primitive meteorites.

### 3.2 Transmission Electron Microscopy

We successfully extracted four FIB sections containing five presolar silicate grains and analyzed them by TEM. Note that the names of the grains have been changed since initial reports of the TEM data (Stroud et al., 2013; Stroud et al., 2014), as indicated below. Compositional results for the five grains are given in Table 4.

*DOM-3 (formerly A2-18):* Scanning TEM (STEM)-based EDS mapping revealed this Group 1 presolar grain to have an Fe-rich (Mg/(Mg+Fe)=0.38), non-stoichiometric composition with an (Fe+Mg+Ca)/Si ratio of 1.5, which is intermediate between those of olivine and pyroxene (Table 4). No lattice fringes were identified in high-resolution TEM (HRTEM) imaging, which indicates an amorphous structure, consistent with the intermediate composition.

*DOM-8 (formerly A2C-15):* A STEM-based EDS measurement indicates that this ~500-nm Group 1 grain has a composition that is very close to stoichiometric olivine with a Mg/(Mg+Fe) ratio of 0.83 (Table 4). We were unfortunately unable to make any structural measurements due to the small volume of sample remaining after the SIMS measurement, and the overlap with adjacent grains and protective C strap. However all previous presolar silicates with olivine compositions for which structural information is available were found to be crystalline olivine (e.g., Messenger et al., 2005; Busemann et al., 2009; Vollmer et al., 2009a). We thus consider it most likely that this grain is indeed forsteritic olivine with Fe contents within the range previously seen for presolar grains.

*DOM-17,18 (formerly A2C2-25a,b):* DOM-17 and DOM-18 were located 4 μm apart in the meteorite section and we were thus able to prepare a single FIB section including both grains (Figure 13). DOM-17 is the most $^{17}$O-rich gain we identified in DOM 08006, though its composition still lies within the Group 1 field (Fig. 6). In contrast, the neighboring DOM-18 grain is the most $^{18}$O-poor grain we identified (Fig. 6) and although its composition lies within the Group 1 field, it may well be a Group 2 grain for which a small amount of normal O from surrounding material diluted its anomalous $^{18}O/^{16}O$ ratio. STEM-based EDS mapping (Table 4) indicates an Fe-rich, non-stoichiometric composition for DOM-17, similar to that of DOM-3, and a Mg-rich olivine composition for DOM-18 (Fig. 14). DOM-17 is amorphous, whereas a selected-area diffraction pattern of grain DOM-18 indexes to crystalline olivine (Fig. 14), in agreement with its chemical composition.

*DOM-77 (formerly A3C-CN1):* STEM-based EDS mapping (Figure 15) shows that grain DOM-77 is compositionally zoned, with a Mg-silicate rim surrounding inner zones of Ca- and Al-rich material. HRTEM imaging reveals that the Mg-silicate region is polycrystalline with lattice fringe spacings consistent with forsteritic olivine. Lattice fringes from the Ca-rich zone index to hibonite, anorthite, or monticellite, but not to grossite, gehlenite, or akermanite. No lattice fringes definitively associated with the Al-rich zone were identified. The olivine at the top surface of the grain, where it is most easily distinguished from adjacent matrix, has a Mg/(Mg+Fe)≈0.93. The high Ca/Al ratio (0.5, Table 4) indicates that hibonite is not the dominant Ca-bearing phase. However, the compositions of the distinct Ca- and Al-rich sub-grains could not be accurately determined because they are smaller than the thickness of the FIB section. Overall, the mineralogy of this composite grain is reminiscent of the larger amoeboid olivine aggregates (AOAs) present in chondrites (Scott and Krot, 2003) and IDPs (Joswiak et al., 2013).

## 4. Discussion

### 4.1. Presolar Grains

*4.1.1. Isotopic compositions*

The O-isotopic ratios observed in the 101 presolar O-rich grains (Figures 6 and 7) span similar ranges to those observed before for presolar oxides and silicates, and fall within the four Group definitions of Nittler et al. (1997). As discussed in the literature, the $^{17}$O-rich, $^{18}$O-poor compositions of the largest population of grains (Group 1) point to an origin in asymptotic giant branch (AGB) stars (Huss et al., 1994; Nittler et al., 1994; Nittler et al., 1997; Nittler et al., 2008; Nittler, 2009). The highly $^{18}$O-depleted Group 2 grains have long been considered to have originated in low-mass (<2 $M_\odot$) AGB stars undergoing a poorly-constrained extra mixing process (Wasserburg et al., 1995; Nollett et al., 2003; Palmerini et al., 2017), but a recent measurement of a key nuclear reaction rate suggests that they may instead have formed around more massive, intermediate-mass AGB stars, 4–8 $M_\odot$ (Lugaro et al., 2017). The $^{18}$O-enriched Group 4 grains most likely formed in Type II supernovae (Choi et al., 1998; Nittler et al., 2008), whereas the Group 3 grains, with sub-solar $^{17}$O/$^{16}$O and normal to depleted $^{18}$O/$^{16}$O ratios, likely formed in either supernovae or low-metallicity AGB stars. A small fraction of presolar grains, with $^{17}$O/$^{16}$O>0.004, may have originated in classical novae (Gyngard et al., 2010; Leitner et al., 2012a), but none of the DOM 08006 grains show such compositions. The implications of the grain isotopic signatures for stellar and galactic evolution have been discussed extensively elsewhere (Nittler et al., 2008; Nittler, 2009) and we do not repeat those discussions here.

The $^{12}C/^{13}C$ ratios of the DOM 08006 presolar SiC grains (Table 3 and Figure 8) lie in the range of the dominant "mainstream" population, which makes up ~90% of all presolar SiC (Zinner, 2014) and are thought to have formed around C-rich AGB stars (e.g., Hoppe et al., 1994; Lugaro et al., 2003). Statistically, all of the DOM 08006 SiC grains are almost certainly mainstream grains, although without additional isotopic data we cannot rule out some belong to the rare X or Z classes (≤2% of SiC) as these grains can also have $^{12}C/^{13}C$ ratios in the observed range.

*4.1.2. Abundances*

The matrix-normalized abundances of O-rich presolar grains in the four analyzed areas are given in Table 1. These were calculated by dividing the total area covered by presolar grains by the total area analyzed in the imaging runs (excluding large cracks, etc). To be conservative, we used the grain sizes corrected for the primary beam diameter (Table 2, Figure 4). We assume that the average density of the grains is similar to that of the matrix in order to convert areal density to mass concentration. Three of the four areas show consistently high abundances of 200–300 ppm, whereas Area 7 shows a lower abundance of 155 ppm. However, as discussed earlier, the primary beam was accidentally slightly defocused for this run, leading to poorer spatial resolution and hence poorer detection efficiency for presolar O-rich grains. We thus calculate the average O-rich presolar grain abundance in DOM 08006 to be 257 ppm by combining the data from Areas 2, 3, and 6 (Table 1). Note that without our correction for the primary beam broadening, the inferred abundance would increase to about 300 ppm.

Essentially all published studies have estimated uncertainties on presolar grain abundances simply from Poisson statistics based on the number of identified grains (often using the tables of Gehrels, 1986). However, this method ignores possible contributions to the error budget from the wide range of grain sizes and the uncertainty in determining individual grain sizes. To investigate this further, we used a Monte Carlo method to estimate the uncertainty in our abundance estimate. We calculated 50,000 simulated grain distributions based on the 90 presolar grains identified in Areas 2, 3, and 6. For each instance, we simulated $N$ grains, where $N$ was taken from a Poisson distribution with a mean value of 90. Then, the size of each of the *N* grains was chosen by selecting a random grain from the DOM 08006 data set, of size *n* pixels, and assigning it *n**' pixels from a Poisson distribution with average value *n*. Finally, the size of each simulated grain was corrected for a 120-nm primary beam size as described above for the real data, and a total abundance calculated. The resulting distribution of simulated abundances is shown in Figure 16. We estimate 1σ and 2σ uncertainties based on the limits enclosing 84.1 % and 97.7 % of the simulations, respectively (Table 1 and Figure 16). For comparison, simply using the Gehrels (1986) tables for 90 grains would imply 1σ uncertainty intervals about our average value of 257 ppm of +30 and -27, compared to the values of +44 and -41 that we estimate from the Monte Carlo simulations. It is thus likely that previous reports of presolar silicate abundances in meteorites have underestimated the statistical uncertainties by ignoring the effect of variable grain sizes.

In Figure 17, we compare our derived matrix-normalized abundance of presolar O-rich grains in DOM 08006 to other primitive meteorites, Antarctic micrometeorites, and IDPs. Abundances are shown with 1σ error bars (as reported by original publications) and plotted against the total analyzed area of each sample. We presume that data acquired from larger areas are more statistically robust. Note that we do not include the recent data of Hoppe et al. (2015, 2017). These studies were made with higher NanoSIMS spatial resolution and hence better sensitivity, and thus cannot be directly compared to the data discussed here, all of which was acquired under similar analytical conditions. Our estimates of the presolar silicate and oxide grain abundance in DOM 08006 are higher than those reported in the other primitive carbonaceous chondrites and Antarctic micrometeorites, but still lower than estimated for IDPs, although the 2σ lower limit is comparable to the value of ~180 ppm seen for several primitive meteorites, and the 2σ upper limit overlaps with the error limit on the average abundance in IDPs. Note that to construct Figure 17 we combined two reported sets of abundances for both QUE 99177 (Floss and Stadermann, 2009a; Nguyen et al., 2010) and MET 00426 (Floss and Stadermann, 2009a; Leitner et al., 2016) to decrease the statistical errors. For both meteorites, the Floss and Stadermann (2009a) study found abundances that were greater than 200 ppm, overlapping with our estimate for DOM 08006, but in the case of QUE 99177, analysis of more than twice as much area by Nguyen et al. (2010) indicated a significantly lower abundance for this meteorite. Moreover, the latter study was performed with the same instrument, protocols and analysis software as the present work and is thus directly comparable. The evidence, therefore, suggests that DOM 08006 has a higher abundance of presolar O-rich grains in its matrix than does QUE 99177, though the estimates overlap at the 2$\sigma$ level.

Haenecour et al. (2018) have recently reported presolar grain surveys in DOM 08006, ALH 77307, and the CO3.0 chondrite LaPaz Icefield (LAP) 031117. They found lower abundances in fine-grained chondrule rims compared to interchondrule matrix, but since our study and almost all others have focused on matrix we only show their matrix abundances in Fig. 17. The DOM 08006 matrix abundance found by Haenecour et al. (2018), 227±32 ppm, agrees very well with ours. We thus combined the two data sets to arrive at a weighted average abundance of O-rich presolar grains in DOM 08006 of 240±30 ppm (Fig. 17). This 1$\sigma$ error is estimated from counting statistics of the total of 141 grains, and multiplied by 1.5, based on the Monte Carlo results described above.

The data shown in Figure 17 provide strong support that the matrix of DOM 08006 has an abundance of presolar silicates and oxides that is higher than any other carbonaceous chondrite studied to date and is thus a highly primitive meteorite. We note that a comparably high abundance of presolar silicates (275±50 ppm, based on 32 grains) has been reported for an ordinary chondrite, Meteorite Hills 00526, in a conference abstract (Floss and Haenecour, 2016a). When we compare our results for DOM 08006 to those of QUE 99177 and ALH 77307, acquired in our laboratory by means of very similar analytical techniques (Nguyen et al., 2010), it is apparent that presolar grains are better preserved in DOM 08006. The ratio of presolar silicate grains to presolar oxide grains has also been taken as a sensitive indicator of alteration in a given meteorite, as silicate minerals are more susceptible to destruction by thermal metamorphism or aqueous alteration than are oxide minerals (Floss and Stadermann, 2009a; Leitner et al., 2012b). As discussed above in Section 3.1.1., our data indicate a lower limit on this ratio of 4.4, but the true ratio is likely much higher and indeed Haenecour et al. (2018) reported a ratio of 24.5 for this meteorite, though their statistical uncertainty is high as their estimate is based on 49 silicate grains and 2 oxides. Leitner et al. (2012b) made a simple estimate based on elemental abundances that the silicate/oxide dust ratio in stardust forming around AGB stars (the dominant sources of the presolar O-rich grains) should be ~23, similar to the average value of 22 they estimated for primitive CR chondrites and IDPs. The high apparent ratio in DOM 08006 is further evidence that this is a highly primitive meteorite.

We calculated the matrix-normalized abundance of SiC in DOM 08006 by the same methodology discussed above for presolar O-rich grains, with the exception that we included Area 7 in the calculation. The slightly poorer spatial resolution of the Area 7 measurements has a smaller effect on the detection efficiency of SiC than of O-rich phases since there is much less C surrounding the grains to dilute the isotopic anomalies. We obtained an average abundance of 35 ppm. Monte Carlo modeling like that shown for O-rich grains in Figure 16 indicates 1$\sigma$ errors of ±10 ppm about the average and 2$\sigma$ errors of +25 and -17 ppm; that is, at a 97.7% certainty level, the abundance lies between 18 ppm and 60 ppm. Davidson et al. (2014a) used NanoSIMS ion imaging of IOM residues to estimate presolar SiC abundances in members of various primitive meteorite groups. They found that CI chondrites, most CR chondrites, and the highly primitive ungrouped Acfer 094 all have a roughly constant abundance of presolar SiC of ~30 ppm. Leitner et al. (2012b) found a higher abundance of 131±53 ppm in the CR2 NWA 852, but this value is within uncertainty of the Davidson et al. (2014a) CR data. In contrast, both in situ NanoSIMS and noble gas data imply a lower abundance of SiC of ~10 ppm in the CO3.0 ALH 77307 (Huss et al., 2003; Davidson et al., 2014a). Davidson et al. (2014a) interpreted the constant SiC abundance of ~30 ppm seen for many primitive meteorites as indicating a uniform abundance in the chondrite-forming region of the solar nebula, with the lower abundances seen in some meteorites (e.g., ALH 77307) reflecting partial grain destruction by parent body processing such as thermal metamorphism. More recently, by combining their data with those of Bose et al. (2012), Haenecour et al. (2018) found a higher abundance of 59±13 ppm for C-anomalous grains (assumed to be mostly SiC) in ALH 77307, consistent within errors with the Davidson et al. (2014a) value of 30 ppm. The source of the discrepancy between this value and the lower ones reported for ALH 77307 by Huss et al. (2003) and Davidson et al. (2014a) is unclear, but may indicate that this meteorite experienced heterogeneous thermal metamorphism on relatively small scales. All in all, our SiC abundance for DOM 08006 is clearly in very good agreement with the average seen in most primitive chondrites, supporting the idea that different chondrite classes accreted matrix with a constant amount of presolar SiC. We note that we found an uneven distribution of SiC grains in the four analyzed areas of the thin section (Table 1), with five to seven identified grains in each of Areas 2, 3, and 7, and zero in Area 6. However, the total imaged area of Area 6 was considerably smaller than the other areas and the non-detection of any SiC grains is statistically consistent with a 35 ppm average abundance.

*4.1.3. Mineralogy of presolar silicates*

Most astronomical knowledge of circumstellar dust mineralogy is based on infrared spectroscopy, which has revealed a range of amorphous and crystalline silicate and oxide phases around O-rich AGB stars (e.g., Molster et al., 2010). However, stellar infrared spectra reflect averages over large numbers of individual grains and are often difficult to unambiguously interpret. In this regard, presolar grains are a complementary and valuable tool for understanding the range of compositions and structures of AGB stardust and thus for better understanding stellar dust formation processes.

As discussed in Section 3.2 we obtained TEM data on fives presolar silicate grains (Figs. 13-15, Table 4). Two grains are amorphous and non-stoichiometric with Fe-rich compositions intermediate between those of pyroxene and olivine, two grains are Mg-rich olivine (tentative for DOM-8), and one grain—DOM-77—is a composite with Al and Ca-rich material surrounded by nanocrystalline forsteritic olivine. TEM data have now been reported for more than 50 presolar silicates from meteorites and IDPs (reviewed in Floss and Haenecour, 2016b). Also, reliable major-element chemical data from Auger spectroscopy have been reported for more than 400 presolar silicate grains (e.g., Floss and Stadermann, 2009a; Stadermann et al., 2009; Nguyen et al., 2010; Bose et al., 2012; Floss and Haenecour, 2016b; Haenecour et al., 2018). The compositions and structures of the four non-composite grains analyzed here are within the range of previous observations of presolar silicates in general and in CO3.0 chondrites specifically. For example, Haenecour et al. (2018) classified presolar silicates on the basis of their (Fe+Mg+Ca)/Si ratios and found that grains with compositions that are olivine-like or intermediate between olivine and pyroxene, like the grains reported here, each make up roughly a quarter of the presolar grains in CO3.0 chondrites. Moreover, the two non-stoichiometric amorphous grains reported here, DOM-3 and DOM-17, are Fe-rich, with atomic Fe/Mg ratios of 1.6 and 2.6 and Fe contents of 14 at.% and 17 at.%, respectively. This Fe-richness is typical of presolar silicates; two-thirds of the grains from CO3.0 chondrites analyzed by Auger have Fe/Si>1.2 and the median Fe contents for CO3.0 presolar grains is 17 at.% (Haenecour et al., 2018). We did not identify any pyroxene-like or $SiO_2$-rich or –poor grains, like those seen in the Auger studies, but our TEM statistics are far too limited to make realistic comparisons with the much larger Auger datasets. Finally, considering the available TEM data for presolar silicates as a whole, roughly two-thirds of the analyzed grains from both chondrites and IDPs are amorphous and the rest are at least partially crystalline. Again, our statistics are too small to make quantitative conclusions, but the observation of both amorphous and crystalline grains in our limited data set appears to be consistent with the crystalline fraction seen in the larger data set. All in all, the four non-composite presolar silicates from DOM 08006 analyzed by TEM are not atypical of the population of presolar silicates as a whole.

In contrast, composite grains like DOM-77 are rare among the presolar silicate population, but the compositions and structures of such grains likely capture changing conditions (e.g., temperature, composition) in the stellar gases from which they condensed and thus in principle can provide clues to grain formation in stellar environments. A few other presolar grains containing Ca-Al-rich phases co-existing with ferromagnesian silicates have been reported, including an aggregate of 30-nm scale Ca or Al oxides interspersed with Mg silicates (Nguyen et al., 2010), an amorphous Ca-Si-rich grain with embedded nanocrystals of hibonite (Vollmer et al., 2013), a $MgAl_2O_4$ spinel surrounded by a nonstoichiometric amorphous Mg-rich silicate (Nguyen et al., 2014), an oxide-silicate aggregate with hibonite and Mg-rich olivine composition (Floss and Stadermann, 2012), and a large Al-Ca-Ti oxide grain mantled by a Ca-rich silicate with a pyroxene-like composition (Leitner et al., 2017). Note that no TEM data are available for the latter two grains, so there are no constraints on the microstructures of the various phases included. Of these, grain DOM-77 is somewhat reminiscent of the hibonite-olivine composite reported by Floss and Stadermann (2012), though that object did not have the core-mantle structure of DOM-77. Moreover, as discussed above in Section 3.2, the core of DOM-77 is too Ca-rich to be hibonite, though the TEM measurements did not fully resolve the substructure of the core.

The composite structure of a grain with a core enriched in refractory elements like Ca and Al, mantled by less refractory Mg-rich silicate, is highly suggestive of condensation from a cooling gas. That is, in a high-temperature (>1300 K) gas of solar composition (a good approximation for O-rich AGB star envelopes) under thermodynamic equilibrium, oxides and silicates rich in Al, Ca, and Ti (e.g., corundum, hibonite, perovskite, grossite, melilite, etc.) condense, followed by Mg-rich silicates (forsterite followed by enstatite), iron metal, and other phases at lower T as the earlier formed phases react with the ambient gas. However, models of circumstellar dust formation in AGB outflows suggest that the process is not fully controlled by equilibrium condensation, with the kinetics of grain nucleation and growth playing a key role (Höfner et al., 2009; Gail, 2010). As direct samples of AGB dust production, presolar grains obviously can provide important clues to the process. TEM studies of refractory Al-rich presolar oxide grains (e.g., corundum, spinel, and hibonite) have revealed most, but not all, to be well-crystallized close-to-stoichiometric minerals that probably formed by equilibrium condensation at high temperature (Stroud et al., 2004; Zega et al., 2011; Takigawa et al., 2014; Zega et al., 2014) and a similar conclusion can be drawn for presolar SiC (Daulton et al., 2003). This suggests that grain condensation at the highest temperatures in AGB stellar envelopes is governed by thermodynamic equilibrium, even though kinetics must play a critical role in the nucleation and growth of circumstellar dust grains (Bernatowicz et al., 1996; Gail and Sedlmayr, 1999). In contrast, the dizzying range of non-stoichiometric compositions and microstructures observed for presolar silicates (Floss and Haenecour, 2016b) and Fe-Cr-rich spinels (Zega et al., 2014) indicates formation far from equilibrium, likely dominated by kinetic effects.

Core-mantle grains like DOM-77 and the others described above provide the opportunity to probe the transition from equilibrium to non-equilibrium condensation in stellar outflows. For example, the mantling of a crystalline spinel grain by amorphous silicate seen by Nguyen et al. (2014) strongly suggests such a transition in the parent star of that presolar grain. Yet the crystalline nature of both the Al-Ca-rich core phases and the Mg-rich olivine rim of DOM-77 suggests that, in contrast, equilibrium conditions prevailed through the full history of the grain's condensation. In fact, the composition of this grain is similar to that of amoeboid olivine aggregates in primitive meteoritic materials (Scott and Krot, 2003; Joswiak et al., 2013), widely considered to be equilibrium condensates from the solar nebula (Grossman and Steele, 1976). A similar argument may be made for the oxide-silicate grain reported by Floss and Stadermann (2012), but the crystallinity of that grain is unknown. A detailed study of grain formation in stellar environments to quantify what conditions may have led to the observed properties of DOM-77 is beyond the scope of this paper, but this result clearly adds to the evidence of highly variable and dynamic conditions during the formation of dust around AGB stars.

## 4.2. Organic matter

The C NanoSIMS images acquired here revealed abundant sub-μm carbonaceous inclusions dispersed throughout the matrix of DOM 08006 (Figure 3; Section 3.1.2), similar to what has been seen in SIMS studies of other carbonaceous chondrites (Nakamura-Messenger et al., 2006; Remusat et al., 2010; Bose et al., 2012; Peeters et al., 2012; Floss et al., 2014; Le Guillou et al., 2014; Alexander et al., 2017). Although it is difficult to ascertain the original morphology of the grains from SEM images, at least one of the $^{15}N$-enriched grains appears to be a nanoglobule (Fig. 3; Garvie and Buseck, 2004; Nakamura-Messenger et al., 2006; Floss and Stadermann, 2009b), providing further evidence that these are a common constituent of organic matter in all classes of primitive extraterrestrial materials including chondrites, micrometeorites, IDPs and comet Wild-2 samples (Dobrică et al., 2009; De Gregorio et al., 2010; Matrajt et al., 2012; De Gregorio et al., 2013). De Gregorio et al. (2013) found evidence for two chemical populations of nanoglobules in meteoritic IOM, one with similar C functional groups to the non-globular IOM and one showing a distinctly more aromatic character. Unfortunately, we do not have functional-group data for nanoglobules in DOM 08006 to make a direct comparison.

The origin of the IOM in extraterrestrial materials is unclear and controversial (Alexander et al., 2017), although the most extreme H and N isotopic anomalies are commonly taken as evidence for an interstellar provenance for at least a portion of the matter (Messenger, 2000; Busemann et al., 2006). This reflects astronomical observations of interstellar molecules showing a wide range of D/H and $^{15}N/^{14}N$ ratios as well as the results of theoretical astrochemical models predicting large fractionations in molecular clouds due to a variety of processes, including for example low-temperature gas-phase ion molecule and gas-grain chemical reactions (Ehrenfreund and Charnley, 2000; Sandford et al., 2001). Carbon isotopic anomalies are much rarer in meteoritic organics, but they are also predicted by astrochemical modeling and interstellar chemistry has thus been invoked to explain these as well (Floss and Stadermann, 2009b; Bose et al., 2014). The observation of similar C- and N-isotopically anomalous sub-μm organic grains in the CO3s DOM 08006 and ALH 77307 (this work; Bose et al., 2014), CRs (Busemann et al., 2006; Floss and Stadermann, 2009b), IDPs of possible cometary origin (Floss et al., 2004) and comet Wild 2 samples (McKeegan et al., 2006; De Gregorio et al., 2010) suggests that such grains were distributed throughout the protosolar disk.

Meteoritic organic matter is chemically and isotopically heterogeneous on all scales, from individual sub-μm grains as studied here, to bulk meteorites. A key unanswered question regarding meteoritic organics is whether differences in bulk organic properties between chondrite groups reflects divergent parent-body processing of a common precursor material or that the organics accreted by different groups was fundamentally different. Data from different portions of the primitive ungrouped carbonaceous chondrite Tagish Lake show that much of the variability in the chemical and isotopic properties of both soluble and insoluble organics in CI, CM and CR chondrites is probably due to aqueous alteration in the meteorites' parent asteroids (Herd et al., 2011; Alexander et al., 2014). The IOM in CR chondrites and the unusual CM chondrite Bells bears many similarities to that found in IDPs and inferred for comet Halley and is probably the most primitive of any meteoritic IOM (Alexander et al., 2007). In contrast, with the possible exception of LL3.00 Semarkona, the IOM in even the most primitive members of chondrite groups that have experienced thermal metamorphism (e.g., unequilibrated ordinary, CO3, and CV3 chondrites) appears to have experienced some heating relative to that found in CRs, based on compositional data, Raman spectroscopy, and X-ray absorption spectroscopy (Bonal et al., 2006; Alexander et al., 2007; Busemann et al., 2007; Cody et al., 2008; Bonal et al., 2016) and its properties can be a sensitive probe of the degree of metamorphism.

Bonal et al. (2016) found that Raman spectral parameters of organic matter in DOM 08006 indicate less heating than that experienced by ALH 77307, previously thought to be the most primitive CO3.0 chondrite. This conclusion is supported by the data of Alexander et al. (2018), who found that DOM 08006 has a higher bulk C abundance, and its IOM has higher H/C, D/H and $^{15}N/^{14}N$ ratios than ALH 77307. Because the H/C ratio of IOM in DOM 08006 is similar to that in the unequilibrated ordinary chondrite (LL3.00) Semarkona, Alexander et al. (2018) suggested that the two meteorites experienced similar thermal histories and that DOM 08006 should thus be classified as a 3.00. As discussed above in Section 3.1.3, our NanoSIMS measurements revealed no resolvable heterogeneity in D/H ratios in DOM 08006, either *in situ* or in isolated IOM, consistent with loss due to metamorphism (Alexander et al., 2010). The lower D/H and $^{15}N/^{14}N$ values for bulk IOM from DOM 08006 compared to primitive CR chondrites indicate that the metamorphism has affected both elements, but the preservation of $^{15}$N-rich hotspots as well as C-anomalous grains in DOM 08006 indicate that the molecular carriers of these anomalies are more thermally resistant than those of D anomalies within primitive IOM. Thus, it appears that D-rich moieties in IOM are less resistant both to thermal metamorphism and to aqueous alteration (Herd et al., 2011) on parent bodies than are carriers of N anomalies. Moreover, the loss of D-rich, and to a lesser extent, $^{15}$N-rich, organic molecules without significantly affecting other inorganic signs of primitiveness (e.g., the high presolar grain abundances reported here; the high Cr contents of olivine reported by Davidson et al., 2014b) provides further evidence that IOM properties are among the most sensitive probes of metamorphism in highly primitive chondrites. In this regard, the increasing recognition that organics can be used as sensitive probes of parent-body processing suggests that a new petrologic scale, including consideration of organic properties, may be needed to better describe the nature of highly primitive chondrites. However, this would require a better understanding than currently exists of the response of organic matter to parent-body processing under a wide range of conditions (see, e.g., the discussion of difficulties arising when attempting to define a Raman-based C thermometer for Type 3 chondrites in Bonal et al., 2016).

Based on the total areas of identified anomalous grains relative to the total mapped areas, and correcting for an assumed factor of two lower density of IOM compared to the matrix as a whole, we estimate abundances of ~33 ppm and ~135 ppm for C-anomalous and N-anomalous grains, respectively; grains anomalous in both elements are counted in both abundance estimates. However, if we limit consideration to the combined C- and N-isotopic imaging runs of matrix Area 3 ("A3cCN" grains in Table 3), we find a much higher abundance of C-anomalous grains (~400 ppm). This higher abundance may indicate true heterogeneity in the meteorite, but more likely reflects differences in image analysis methodology (automatic definition of C-rich regions of interest as opposed to ratio image generation). Thus, the data are consistent with spatial heterogeneity in the abundance of C-anomalous grains, but additional systematic C and N mapping would be required to tell this definitively. In comparison, Floss and Stadermann (2009b) found an abundance of C-anomalous grains in the CR chondrites QUE 99177 and MET 00426 of ~120 ppm and Bose et al. (2012) reported an abundance of $^{15}$N-rich grains in CO3.0 ALH 77307 of 160±30 ppm. Neither study provided details of how the abundances were estimated (e.g., whether or not a density correction was applied to convert area fraction to mass fraction). Nevertheless, it is apparent that the abundances of C-anomalous and $^{15}$N-rich grains in DOM 08006 are roughly similar to those in the least altered CR chondrites, providing further support that this is a highly primitive meteorite.

Note that Bose et al. (2014) reported a remarkably high abundance (~900 ppm) of very $^{15}$N-rich grains, with $\delta^{15}$N values ranging from 1600 to 3000, in the CO3 chondrite QUE 97416. These authors further identified a single presolar silicate grain, with an inferred presolar silicate abundance of 6 ppm, indicating that this meteorite is likely more processed than either DOM 08006 or ALH 77307. Given this, the high abundance of extremely $^{15}$N-rich grains is puzzling; similar grains were not identified here and are absent or very rare in other chondrites (e.g., Figure 10). Possible explanations are that QUE 97416 originated on a different parent body than other CO3s, one that sampled a different population of anomalous organic grains, that thermal metamorphism or some other process somehow enriched some organic grains in $^{15}$N in QUE 97416, or that the data reported by Bose et al. (2014) are compromised by an unrecognized instrumental artifact.

As mentioned above in Section 3.1.2, the NanoSIMS mapping revealed a grain, A7_60 (Figures 10 and 11), with large depletions in both $^{13}$C and $^{15}$N relative to terrestrial isotopic compositions ($\delta^{13}$C=-300 ‰, $^{12}$C/$^{13}$C=127, $\delta^{15}$N=-264 ‰, $^{14}$N/$^{15}$N=371). These isotopic compositions are similar to those of some rare presolar SiC and graphite grains thought to come from low-metallicity AGB stars (e.g., Figs. 3 and 12 of Zinner, 2014); SEM-EDS analysis only revealed C for this grain raising the possibility that it is indeed a presolar graphite that got sheared into two fragments during polishing of the DOM 08006 thin section. However, the NanoSIMS measurement of this grain revealed it to be rich in N, with a measured $CN^-/C^-$ secondary ion ratio of ~0.8, roughly corresponding to ~10% N (Alleon et al., 2015), suggesting that it is in fact N-rich organic matter and not graphite. This does not rule out it being a presolar circumstellar organic grain, as organic molecules are indeed observed around evolved stars (Matsuura et al., 2014), but it does have a relatively unusual isotopic composition for AGB stardust. Alternatively, the N-isotopic composition of grain A7_60 is close to, but not as $^{14}$N-rich as, that of the solar wind ($\delta^{15}$N ≈ -400 ‰; Marty et al., 2011), which is assumed to be representative of the bulk protosolar composition. The C-isotopic composition of the Sun is unknown. Hashizume et al. (2004) argued on the basis of SIMS depth profile measurements of lunar soils that it is isotopically light relative to the Earth, with $\delta^{13}$C <-100 ‰. Recently, Lyons et al. (2017) reported a spectroscopic measurement of CO in the solar photosphere, yielding $\delta^{13}$C = -48 ± 7‰. Thus, although the lunar measurement of Hashizume et al. (2004) does not rule out as extreme a $^{13}C$ depletion in the Sun as observed in A7_60, the astronomical measurement argues against it. We suggest that the most likely origin of grain A7_60 is in the Sun's parental molecular cloud, where its C isotopic composition was set by interstellar fractionation processes that were decoupled from those affecting N isotopes (e.g., Langer et al., 1984), with the latter mainly reflecting the bulk isotopic composition of the cloud.

## 5. Conclusions

A detailed NanoSIMS and TEM study of a polished thin section of the CO3.0 chondrite DOM 08006 allows us to draw the following conclusions:

1) NanoSIMS isotopic mapping revealed that DOM 08006 has the highest abundance of presolar O-rich (silicate and oxide) grains in its matrix, 257 +76/-96 ppm (2σ), found thus far in any carbonaceous chondrite, in good agreement with an independent measurement of a different thin section by Haenecour et al. (2018). Combining our dataset with that of Haenecour et al. (2018) gives an average abundance of 240±30 ppm (1σ). The NanoSIMS data also revealed a SiC abundance of 35 (+25/-17, 2σ) ppm, similar to that seen in other highly primitive meteorites. Within the statistical uncertainty of our observations, the abundance of presolar grains in DOM 08006 is homogeneous on a scale of several mm. The distributions of isotopic compositions of the presolar grains identified here are consistent with previous observations in other primitive meteorites with differences being fully explainable by analytical effects (e.g., dilution of isotopic signatures by surrounding material during NanoSIMS measurements).
2) Transmission electron microscopy was performed on five presolar silicates. One grain was found to have a composite mineralogy that is similar to larger amoeboid olivine aggregates found in chondrites, and consistent with condensation near thermodynamic equilibrium. In addition, the TEM observations revealed two non-stoichiometric amorphous grains and two olivine grains, though one of the latter is identified as such solely based on its composition.
3) NanoSIMS C and N mapping revealed IOM to be present primarily as sub-micron inclusions with ranges of C- and N-isotopic anomalies similar to those seen in primitive

CR chondrites and interplanetary dust particles, but with normal (terrestrial) and homogeneous D/H ratios. Most likely, DOM 08006 and other CO chondrites accreted a similar complement of primitive and isotopically anomalous organic matter to that found in other chondrite classes and IDPs, but the very limited amount of thermal metamorphism experienced by DOM 08006 has caused loss of D-rich organic moieties, while not substantially affecting the molecular carriers of C and N anomalies or the inorganic phases of the meteorite. In this regard, the D/H of organic matter may be the most sensitive probe available of thermal metamorphism in highly primitive chondrites. One organic grain that was highly depleted in $^{13}C$ and $^{15}N$ was identified; we propose it originated in the Sun's parental molecular cloud.

4) Together with other isotopic and petrographic evidence (Davidson et al., 2014b; Bonal et al., 2016; Alexander et al., 2018), our data indicate that DOM 08006 has experienced the least parent body modification of any known CO chondrite. Compared to other highly primitive carbonaceous chondrites (e.g., CRs QUE 99177 and MET 00426, and the ungrouped Acfer 094), it contains a higher abundance of O-rich presolar grains in its matrix as well as organic matter with similar ranges of C and N isotopic anomalies, but more homogeneous D/H ratios. These differences most likely reflect the effects of the limited parent body processing under variable conditions that these meteorites experienced (e.g., aqueous alteration for CRs and Acfer 094, heating for DOM 08006). The fact that no uniquely primitive single meteorite has yet been identified with no signs of parent body processing highlights the need for continued comparative work on all of these samples to help further elucidate the earliest stages of parent body processing on volatile-rich asteroids.

***Acknowledgements:*** The authors would like to thank Cecilia Satterwhite and Kevin Righter (NASA, Johnson Space Center) for supplying the section of DOM 08006. US Antarctic meteorite samples are recovered by the Antarctic Search for Meteorites (ANSMET) program, which has been funded by NSF and NASA, and characterized and curated by the Department of Mineral Sciences of the Smithsonian Institution and the Astromaterials Curation Office at NASA Johnson Space Center. We thank the Associate Editor, Eric Quirico, and the referees, Christine Floss and Henner Busemann for constructive and helpful comments. We also acknowledge Ryan Ogliore for inspiring the Monte Carlo error estimation method used in this paper. This work was supported in part by NASA grants NNX10AI63G and NNX11AB40G.

## References

Alexander, C.M.O'D., Cody, G.D., Gregorio, B.T.D., Nittler, L.R. and Stroud, R.M. (2017) The nature, origin and modification of insoluble organic matter in chondrites, the major source of Earth's C and N. *Chemie der Erde - Geochemistry* **77**, 227-256.

Alexander, C.M.O'D., Cody, G.D., Kebukawa, Y., Bowden, R., Fogel, M.L., Kilcoyne, A.L.D., Nittler, L.R. and Herd, C.D.K. (2014) Elemental, isotopic, and structural changes in Tagish Lake insoluble organic matter produced by parent body processes. *Meteorit. Planet. Sci.* **49**, 503–525.

Alexander, C.M.O'D., Fogel, M., Yabuta, H. and Cody, G.D. (2007) The origin and evolution of chondrites recorded in the elemental and isotopic compositions of their macromolecular organic matter. *Geochim. Cosmochim. Acta* **71**, 4380–4403.

Alexander, C.M.O'D., Greenwood, R.C., Bowden, R., Gibson, J.M., Howard, K.T. and Franchi, I.A. (2018) A multi-technique search for the most primitive CO chondrites. *Geochim. Cosmochim. Acta* **221,** 406-420.

Alexander, C.M.O'D., Howard, K.T., Bowden, R. and Fogel, M.L. (2013) The classification of CM and CR chondrites using bulk H, C and N abundances and isotopic compositions. *Geochim. Cosmochim. Acta* **123**, 244–260.

Alexander, C.M.O'D., Newsome, S.D., Fogel, M.L., Nittler, L.R., Busemann, H. and Cody, G.D. (2010) Deuterium enrichments in chondritic macromolecular material--Implications for the origin and evolution of organics, water and asteroids. *Geochim. Cosmochim. Acta* **74**, 4417–4437.

Alexander, C.M.O'D., Nittler, L.R., Davidson, J. and Ciesla, F.J. (2017c) Measuring the level of interstellar inheritance in the solar protoplanetary disk. *Meteorit. Planet. Sci.* **52**, 1797-1821.

Alleon, J., Bernard, S., Remusat, L. and Robert, F. (2015) Estimation of nitrogen-to-carbon ratios of organics and carbon materials at the submicrometer scale. *Carbon* **84**, 290-298.

Bernatowicz, T.J., Cowsik, R., Gibbons, P.C., Lodders, K., Fegley, B., Jr., Amari, S. and Lewis, R.S. (1996) Constraints on stellar grain formation from presolar graphite in the Murchison meteorite. *Ap. J.* **472**, 760–782.

Bonal, L., Quirico, E., Bourot-Denise, M. and Montagnac, G. (2006) Determination of the petrologic type of CV3 chondrites by Raman spectroscopy of included organic matter. *Geochim. Cosmochim. Acta* **70**, 1849.

Bonal, L., Quirico, E., Flandinet, L. and Montagnac, G. (2016) Thermal history of type 3 chondrites from the Antarctic meteorite collection determined by Raman spectroscopy of their polyaromatic carbonaceous matter. *Geochim. Cosmochim. Acta* **189**, 312-337.

Bose, M., Floss, C., Stadermann, F.J., Stroud, R.M. and Speck, A.K. (2012) Circumstellar and interstellar material in the CO3 chondrite ALHA77307: An isotopic and elemental investigation. *Geochim. Cosmochim. Acta* **93**, 77–101.

Bose, M., Zega, T.J. and Williams, P. (2014) Assessment of alteration processes on circumstellar and interstellar grains in Queen Alexandra Range 97416. *Earth Planet. Sci. Lett.* **399**, 128–138.

Brearley, A.J. (1993) Matrix and fine-grained rims in the unequilibrated CO3 chondrite, ALHA 77307: Origins and evidence for diverse, primitive nebular dust components. *Geochim. Cosmochim. Acta* **57**, 1521–1550.

Burton, A.S., Elsila, J.E., Callahan, M.P., Martin, M.G., Glavin, D.P., Johnson, N.M. and Dworkin, J.P. (2012) A propensity for n-omega-amino acids in thermally altered Antarctic meteorites. *Meteorit. Planet. Sci.* **47**, 374–386.

Busemann, H., Alexander, C.M.O'D. and Nittler, L.R. (2007) Characterization of insoluble organic matter in meteorites by Raman spectroscopy. *Meteorit. Planet. Sci.* **42**, 1387–1416.

Busemann, H., Nguyen, A.N., Cody, G.D., Hoppe, P., Kilcoyne, A.L.D., Stroud, R.M., Zega, T.J. and Nittler, L.R. (2009) Ultra-primitive interplanetary dust particles from the comet 26P/Grigg-Skjellerup dust stream collection. *Earth Planet. Sci. Lett.* **288**, 44–57.

Busemann, H., Young, A.F., Alexander, C.M.O'D., Hoppe, P., Mukhopadhyay, S. and Nittler, L.R. (2006) Interstellar chemistry recorded in organic matter from primitive meteorites. *Science* **312**, 727–730.

Choi, B.-G., Huss, G.R. and Wasserburg, G.J. (1998) Presolar corundum and spinel in ordinary chondrites: Origins from AGB stars and a supernova. *Science* **282**, 1282–1289.

Ciesla, F.J. and Sandford, S.A. (2012) Organic synthesis via irradiation and warming of ice grains in the solar nebula. *Science* **336**, 452–454.

Cody, G.D., Alexander, C.M.O'D., Yabuta, H., Kilcoyne, A.L.D., Araki, T., Ade, H., Dera, P., Fogel, M., Militzer, B. and Mysen, B.O. (2008) Organic thermometry for chondritic parent bodies. *Earth Planet. Sci. Lett.* **272**, 446–455.

Cody, G.D., Heying, E., Alexander, C.M.O'D., Nittler, L.R., Kilcoyne, A.L.D., Sandford, S.A. and Stroud, R.M. (2011) Establishing a molecular relationship between chondritic and cometary organic solids. *Proc. Nat. Acad. Sci. USA* **108**, 19171–19176

Dai, Z.R., Bradley, J.P., Joswiak, D.J., Brownlee, D.E., Hill, H.G.M. and Genge, M.J. (2002) Possible in situ formation of meteoritic nanodiamonds in the early Solar System. *Nature* **418**, 157–159.

Daulton, T.L., Bernatowicz, T.J., Lewis, R.S., Messenger, S., Stadermann, F.J. and Amari, S. (2003) Polytype distribution of circumstellar silicon carbide - microstructural characterization by transmission electron microscopy. *Geochim. Cosmochim. Acta* **67**, 4743–4767.

Davidson, J., Busemann, H. and Franchi, I.A. (2012) A NanoSIMS and Raman spectroscopic comparison of interplanetary dust particles from comet Grigg-Skjellerup and non-Grigg Skjellerup collections. *Meteorit. Planet. Sci.* **47**, 1748–1771.

Davidson, J., Busemann, H., Nittler, L.R., Alexander, C.M.O'D., Orthous-Daunay, F.-R., Franchi, I.A. and Hoppe, P. (2014a) Abundances of presolar silicon carbide grains in primitive meteorites determined by NanoSIMS. *Geochim. Cosmochim. Acta* **139**, 248–266.

Davidson, J., Nittler, L.R., Alexander, C.M.O'D. and Stroud, R.M. (2014b) Petrography of very primitive CO3 chondrites: Dominion Range 08006, Miller Range 07687, and four others. Lunar Planet. Sci. XLV, #1384 (abstr.).

De Gregorio, B.T., Stroud, R.M., Nittler, L.R., Alexander, C.M.O'D., Bassim, N.D., Cody, G.D., Kilcoyne, A.L.D., Sandford, S.A., Milam, S.N., Nuevo, M. and Zega, T.J. (2013) Isotopic and chemical variation of organic nanoglobules in primitive meteorites. *Meteorit. Planet. Sci.* **48**, 804–828.

De Gregorio, B.T., Stroud, R.M., Nittler, L.R., Alexander, C.M.O'D., Kilcoyne, A.L.D. and Zega, T.J. (2010) Isotopic anomalies in organic nanoglobules from Comet 81P/Wild 2: Comparison to Murchison nanoglobules and isotopic anomalies induced in terrestrial organics by electron irradiation. *Geochim. Cosmochim. Acta* **74**, 4454–4470.

Dobrică, E., Engrand, C., Duprat, J., Gounelle, M., Leroux, H., Quirico, E. and Rouzaud, J.-N. (2009) Connection between micrometeorites and Wild 2 particles: From Antarctic snow to cometary ices. *Meteorit. Planet. Sci.* **44**, 1643–1661.

Ehrenfreund, P. and Charnley, S.B. (2000) Organic molecules in the interstellar medium, comet and meteorites: A voyage from dark clouds to the early Earth. *Astron. Astrophys.* **38**, 427–483.

Floss, C. and Haenecour, P. (2016a) Meteorite Hills (MET) 00526: An unequilibrated ordinary chondrite with high presolar grain abundances. Lunar Planet. Sci. XLVII, #1030 (abstr.).

Floss, C. and Haenecour, P. (2016b) Presolar silicate grains: Abundances, isotopic and elemental compositions, and the effects of secondary processing. *Geochemical Journal* **50**, 3-25.

Floss, C., Le Guillou, C. and Brearley, A. (2014) Coordinated NanoSIMS and FIB-TEM analyses of organic matter and associated matrix materials in CR3 chondrites. *Geochim. Cosmochim. Acta* **139**, 1–25.

Floss, C. and Stadermann, F. (2009a) Auger Nanoprobe analysis of presolar ferromagnesian silicate grains from primitive CR chondrites QUE 99177 and MET 00426. *Geochim. Cosmochim. Acta* **73**, 2415–2440.

Floss, C. and Stadermann, F.J. (2009b) High abundances of circumstellar and interstellar C-anomalous phases in the primitive CR3 chondrites QUE 99177 and MET 00426. *Ap. J.* **697**, 1242–1255.

Floss, C. and Stadermann, F.J. (2012) Presolar silicate and oxide abundances and compositions in the ungrouped carbonaceous chondrite Adelaide and the K chondrite Kakangari: The effects of secondary processing. *Meteorit. Planet. Sci.* **47**, 992–1009.

Floss, C., Stadermann, F.J., Bradley, J., Dai, Z.R., Bajt, S. and Graham, G. (2004) Carbon and nitrogen isotopic anomalies in an anhydrous interplanetary dust particle. *Science* **303**, 1355–1358.

Floss, C., Stadermann, F.J., Bradley, J.P., Dai, Z.R., Bajt, S., Graham, G. and Lea, A.S. (2006) Identification of isotopically primitive interplanetary dust particles: A NanoSIMS isotopic imaging study. *Geochim. Cosmochim. Acta* **70**, 2371.

Gail, H.-P. (2010) Formation and evolution of minerals in accretion disks and stellar outflows, in: Henning, T. (Ed.), Astromineralogy. Springer Berlin Heidelberg, Berlin, Heidelberg, pp. 61-141.

Gail, H.-P. and Sedlmayr, E. (1999) Mineral formation in stellar winds. I. Condensation sequence of silicate and iron grains in stationary oxygen rich outflows. *Astron. Astrophys.* **347**, 594–616.

Garvie, L.A.J. and Buseck, P.R. (2004) Nanosized carbon-rich grains in carbonaceous chondrite meteorites. *Earth Planet. Sci. Lett.* **224**, 431–439.

Gehrels, N. (1986) Confidence limits for small numbers of events in astrophysical data. *Ap. J.* **303**, 336–346.

Gourier, D., Robert, F., Delpoux, O., Binet, L., Vezin, H., Moissette, A. and Derenne, S. (2008) Extreme deuterium enrichment of organic radicals in the Orgueil meteorite: Revisiting the interstellar interpretation? *Geochim. Cosmochim. Acta* **72**, 1914–1923.

Greshake, A. (1997) The primitive matrix components of the unique carbonaceous chondrite Acfer 094; A TEM study. *Geochim. Cosmochim. Acta* **61**, 437–452.

Grossman, J.N. and Brearley, A.J. (2005) The onset of metamorphism in ordinary and carbonaceous chondrites. *Meteorit. Planet. Sci.* **40**, 87-122.

Grossman, L. and Steele, I.M. (1976) Amoeboid olivine aggregates in the Allende meteorite. *Geochim. Cosmochim. Acta* **40**, 149-155.

Gyngard, F., Nittler, L., Zinner, E. and Jose, J. (2010) Oxygen rich stardust grains from novae, Nuclei in the Cosmos, p. 141.

Haenecour, P., Floss, C., José, J., Amari, S., Lodders, K., Jadhav, M., Wang, A. and Gyngard, F. (2016) Coordinated analysis of two graphite grains from the CO3.0 LAP 031117 meteorite: First identification of a CO nova graphite and a presolar iron sulfide subgrain. *Ap. J.* **825:** 88.

Haenecour, P., Floss, C., Zega, T.J., Croat, T.K., Wang, A., Jolliff, B.L. and Carpenter, P. (2018) Presolar silicates in the matrix and fine-grained rims around chondrules in primitive CO3.0 chondrites: Evidence for pre-accretionary aqueous alteration of the rims in the solar nebula. *Geochim. Cosmochim. Acta* **221,** 379-405.

Harju, E.R., Rubin, A.E., Ahn, I., Choi, B.-G., Ziegler, K. and Wasson, J.T. (2014) Progressive aqueous alteration of CR carbonaceous chondrites. *Geochim. Cosmochim. Acta* **139**, 267–292.

Hashizume, K., Chaussidon, M., Marty, B. and Terada, K. (2004) Protosolar carbon isotopic composition: Implications for the origin of meteoritic organics. *Ap. J.* **600**, 480–484.

Heck, P.R., Stadermann, F.J., Isheim, D., Auciello, O., Daulton, T.L., Davis, A.M., Elam, J.W., Floss, C., Hiller, J., Larson, D.J., Lewis, J.B., Mane, A., Pellin, M.J., Savina, M.R., Seidman, D.N. and Stephan, T. (2014) Atom-probe analyses of nanodiamonds from Allende. *Meteorit. Planet. Sci.* **49**, 453–467.

Herd, C.D.K., Blinova, A., Simkus, D.N., Huang, Y., Tarozo, R., Alexander, C.M.O'D., Gyngard, F., Nittler, L.R., Cody, G.D., Fogel, M.L., Kebukawa, Y., Kilcoyne, A.L.D., Hilts, R.W., Slater, G.F., Glavin, D.P., Dworkin, J.P., Callahan, M.P., Elsila, J.E., De Gregorio, B.T. and Stroud, R.M. (2011) Origin and evolution of prebiotic organic matter as inferred from the Tagish Lake meteorite. *Science* **332**, 1304–1307.

Höfner, S., Grün, E. and Steinacker, J. (2009) Dust formation and winds around evolved stars: The good, the bad and the ugly cases, in: Henning, T., Eberhard Grün, and Jürgen Steinacker (Eds.), Cosmic Dust - Near and Far, ASP Conference Series, Vol. 414, San Francisco: Astromomical Society of the Pacific, 2009, pp. 3-21.

Hoppe, P., Amari, S., Zinner, E., Ireland, T. and Lewis, R.S. (1994) Carbon, nitrogen, magnesium, silicon, and titanium isotopic compositions of single interstellar silicon carbide grains from the Murchison carbonaceous chondrite. *Ap. J.* **430**, 870–890.

Hoppe, P., Leitner, J. and Kodolányi, J. (2015) New constraints on the abundances of silicate and oxide stardust from supernovae in the Acfer 094 meteorite. *Ap. J.* **808**.

Huss, G.R., Fahey, A.J., Gallino, R. and Wasserburg, G.J. (1994) Oxygen isotopes in circumstellar $Al_2O_3$ grains from meteorites and stellar nucleosynthesis. *Ap. J.* **430**, L81–84.

Huss, G.R. and Lewis, R.S. (1995) Presolar diamond, SiC, and graphite in primitive chondrites: Abundances as a function of meteorite class and petrologic type. *Geochim. Cosmochim. Acta* **59**, 115–160.

Huss, G.R., Meshik, A.P., Smith, J.B. and Hohenberg, C.M. (2003) Presolar diamond, silicon carbide, and graphite in carbonaceous chondrites: implications for thermal processing in the solar nebula. *Geochim. Cosmochim. Acta* **67**, 4823–4848.

Joswiak, D.J., Brownlee, D.E. and Matrajt, G. (2013) First occurrence of a probable amoeboid olivine aggregate in a "cometary" interplanetary dust particle. Lunar Planet. Sci. XLIV, #2410 (abstr.).

Keller, L.P., Messenger, S., Flynn, G.J., Clemett, S., Wirick, S. and Jacobsen, C. (2004) The nature of molecular cloud material in interplanetary dust. *Geochim. Cosmochim. Acta* **68**, 2577-2589.

Langer, W.D., Graedel, T.E., Frerking, M.A. and Armentrout, P.B. (1984) Carbon and oxygen isotope fractionation in dense interstellar clouds. *Ap. J.* **277**, 581-590.

Le Guillou, C., Bernard, S., Brearley, A.J. and Remusat, L. (2014) Evolution of organic matter in Orgueil, Murchison and Renazzo during parent body aqueous alteration: In situ investigations. *Geochim. Cosmochim. Acta* **131**, 368–392.

Le Guillou, C. and Brearley, A. (2014) Relationships between organics, water and early stages of aqueous alteration in the pristine CR3.0 chondrite MET 00426. *Geochim. Cosmochim. Acta* **131**, 344–367.

Leitner, J., Hoppe, P., Floss, C., Hillion, F. and Henkel, T. (2018) Correlated nanoscale characterization of a unique complex oxygen-rich stardust grain: Implications for circumstellar dust formation. *Geochim. Cosmochim. Acta* **221**, 255-274.

Leitner, J., Kodolányi, J., Hoppe, P. and Floss, C. (2012a) Laboratory analysis of presolar silicate stardust from a nova. *Ap. J.* **754**, L41.

Leitner, J., Vollmer, C., Floss, C., Zipfel, J. and Hoppe, P. (2016) Ancient stardust in fine-grained chondrule dust rims from carbonaceous chondrites. *Earth Planet. Sci. Lett.* **434**, 117–128.

Leitner, J., Vollmer, C., Hoppe, P. and Zipfel, J. (2012b) Characterization of presolar material in the CR chondrite Northwest Africa 852. *Ap. J.* **745**, 38.

Lugaro, M., Davis, A.M., Gallino, R., Pellin, M.J., Straniero, O. and Käppeler, F. (2003) Isotopic compositions of strontium, zirconium, molybdenum, and barium in single presolar SiC grains and asymptotic giant branch stars. *Ap. J.* **593**, 486–508.

Lugaro, M., Karakas, A.I., Bruno, C.G., Aliotta, M., Nittler, L.R., Bemmerer, D., Best, A., Boeltzig, A., Broggini, C., Caciolli, A., Cavanna, F., Ciani, G.F., Corvisiero, P., Davinson, T., Depalo, R., Di Leva, A., Elekes, Z., Ferraro, F., Formicola, A., Fülöp, Z., Gervino, G., Guglielmetti, A., Gustavino, C., Gyürky, G., Imbriani, G., Junker, M., Menegazzo, R., Mossa, V., Pantaleo, F.R., Piatti, D., Prati, P., Scott, D.A., Straniero, O., Strieder, F., Szücs, T., Takács, M.P. and Trezzi, D. (2017) Origin of meteoritic stardust unveiled by a revised proton-capture rate of $^{17}$O. *Nature Astronomy* **1**, 0027.

Lyons, J.R., Gharib-Nezhad, E. and Ayres, T.R. (2017) The carbon isotope composition of the Sun. Lunar Planet. Sci. XLVIII, #2309 (abstr.).

Marty, B., Chaussidon, M., Wiens, R.C., Jurewicz, A.J.G. and Burnett, D.S. (2011) A $^{15}$N-poor isotopic composition for the Solar System as shown by Genesis solar wind samples. *Science* **332**, 1533.

Matrajt, G., Messenger, S., Brownlee, D. and Joswiak, D. (2012) Diverse forms of primordial organic matter identified in interplanetary dust particles. *Meteorit. Planet. Sci.* **47**, 525–549.

Matsuura, M., Bernard-Salas, J., Lloyd Evans, T., Volk, K.M., Hrivnak, B.J., Sloan, G.C., Chu, Y.-H., Gruendl, R., Kraemer, K.E., Peeters, E., Szczerba, R., Wood, P.R., Zijlstra, A.A., Hony, S., Ita, Y., Kamath, D., Lagadec, E., Parker, Q.A., Reid, W.A., Shimonishi, T., Van Winckel, H., Woods, P.M., Kemper, F., Meixner, M., Otsuka, M., Sahai, R., Sargent, B.A., Hora, J.L. and McDonald, I. (2014) Spitzer Space Telescope spectra of post-AGB stars in the Large Magellanic Cloud – polycyclic aromatic hydrocarbons at low metallicities. *MNRAS* **439**, 1472-1493.

McKeegan, K.D., Aleon, J., Bradley, J., Brownlee, D., Busemann, H., Butterworth, A., Chaussidon, M., Fallon, S., Floss, C., Gilmour, J., Gounelle, M., Graham, G., Guan, Y., Heck, P.R., Hoppe, P., Hutcheon, I.D., Huth, J., Ishii, H., Ito, M., Jacobsen, S.B., Kearsley, A.,

Leshin, L.A., Liu, M.-C., Lyon, I., Marhas, K., Marty, B., Matrajt, G., Meibom, A., Messenger, S., Mostefaoui, S., Mukhopadhyay, S., Nakamura-Messenger, K., Nittler, L., Palma, R., Pepin, R.O., Papanastassiou, D.A., Robert, F., Schlutter, D., Snead, C.J., Stadermann, F.J., Stroud, R., Tsou, P., Westphal, A., Young, E.D., Ziegler, K., Zimmermann, L. and Zinner, E. (2006) Isotopic compositions of cometary matter returned by Stardust. *Science* **314**, 1724–1728.

Messenger, S. (2000) Identification of molecular cloud material in interplanetary dust particles. *Nature* **404**, 968–971.

Messenger, S., Keller, L.P. and Lauretta, D.S. (2005) Supernova Olivine from Cometary Dust. *Science* **309**, 737–741.

Molster F., Waters L. and Kemper F. (2010) The Mineralogy of Interstellar and Circumstellar Dust in Galaxies. In: Henning T. (Ed.) Astromineralogy. Lecture Notes in Physics, vol 815. Springer, Berlin, Heidelberg, pp. 142-201.

Nakamura-Messenger, K., Messenger, S., Keller, L.P., Clemett, S.J. and Zolensky, M.E. (2006) Organic globules in the Tagish Lake meteorite: Remnants of the protosolar disk. *Science* **314**, 1439–1442.

Nguyen, A.N., Nakamura-Messenger, K., Messenger, S., Keller, L.P. and Klöck, W. (2014) Identification of a compound spinel and silicate presolar grain in a chondritic interplanetary dust particle. Lunar Planet. Sci. XLV, #2351 (abstr.).

Nguyen, A.N., Nittler, L.R., Stadermann, F.J., Stroud, R.M. and Alexander, C.M.O'D. (2010) Coordinated analyses of presolar grains in the Allan Hills 77307 and Queen Elizabeth Range 99177 meteorites. *Ap. J.* **719**, 166–189.

Nguyen, A.N., Stadermann, F.J., Zinner, E., Stroud, R.M., Alexander, C.M.O'D. and Nittler, L.R. (2007) Characterization of presolar silicate and oxide grains in primitive carbonaceous chondrites. *Ap. J.* **656**, 1223–1240.

Nittler, L., Alexander, C.M.O'D., Gao, X., Walker, R.M. and Zinner, E. (1994) Interstellar oxide grains from the Tieschitz ordinary chondrite. *Nature* **370**, 443–446.

Nittler, L.R. (2009) On the mass and metallicity distributions of the parent AGB stars of O-rich presolar stardust grains. *Pub. Astron. Soc. Aust.* **26**, 271–277.

Nittler, L.R., Alexander, C.M.O'D., Gallino, R., Hoppe, P., Nguyen, A., Stadermann, F. and Zinner, E.K. (2008) Aluminum-, calcium- and titanium-rich oxide stardust in ordinary chondrite meteorites. *Ap. J.* **682**, 1450–1478.

Nittler, L.R., Alexander, C.M.O'D., Gao, X., Walker, R.M. and Zinner, E. (1997) Stellar sapphires: The properties and origins of presolar $Al_2O_3$ in meteorites. *Ap. J.* **483**, 475–495.

Nittler, L.R. and Ciesla, F. (2016) Astrophysics with extraterrestrial materials. *Ann. Rev. Astron. Astrophys.* **54**, 53–93.

Nollett, K.M., Busso, M. and Wasserburg, G.J. (2003) Cool bottom processes on the thermally-pulsing AGB and the isotopic composition of circumstellar dust grains. *Ap. J.* **582**, 1036–1058.

Nuth, J.A., III, Johnson, N.M. and Manning, S. (2008) A self-perpetuating catalyst for the production of complex organic molecules in protostellar nebulae. *Ap. J.* **673**, L225–L228.

Palmerini, S., Trippella, O. and Busso, M. (2017) A deep mixing solution to the aluminum and oxygen Isotope puzzles in presolar grains. *MNRAS* **467**, 1193-1201.

Peeters, Z., Changela, H.G., Stroud, R.M., Alexander, C.M.O'D. and Nittler, L.R. (2012) Coordinated analysis of in situ organic material in the CR chondrite QUE 99177. Lunar Planet. Sci. XLIII, #1659 (abstr.).

Pendleton, Y.J., Sandford, S.A., Allamandola, L.J., Tielens, A.G.G.M. and Sellgren, K. (1994) Near-infrared absorption spectroscopy of interstellar hydrocarbon grains. *Ap. J.* **437**, 683–696.

Quirico, E., Raynal, P.I. and Bourot-Denise, M. (2003) Metamorphic grade of organic matter in six unequilibrated ordinary chondrites. *Meteorit. Planet. Sci.* **38**, 795–811.

Remusat, L., Guan, Y., Wang, Y. and Eiler, J.M. (2010) Accretion and preservation of D-rich organic particles in carbonaceous chondrites: Evidence for important transport in the early solar system nebula. *Ap. J.* **713**, 1048–1058.

Robert, F. and Epstein, S. (1982) The concentration and isotopic composition of hydrogen, carbon and nitrogen in carbonaceous meteorites. *Geochim. Cosmochim. Acta* **46**, 81–95.

Sandford, S.A., Bernstein, M.P. and Dworkin, J.P. (2001) Assessment of the interstellar processes leading to deuterium enrichment in meteoritic organics. *Meteorit. Planet. Sci.* **36**, 1117–1133.

Scott, E.R.D. and Krot, A.N. (2003) Chondrites and their Components, in: Davis, A.M. (Ed.), Meteorites, Comets and Planets (Vol. 1), Treatise on Geochemistry (eds: H. D. Holland and K. K. Turekian). Elsevier-Pergamon, Oxford, pp. 143-200.

Slodzian, G., Hillion, F., Stadermann, F.J. and Zinner, E. (2004) QSA influences on isotopic ratio measurements. *Applied Surface Science* **231–232**, 874–877.

Stadermann, F.J., Floss, C., Bose, M. and Lea, A.S. (2009) The use of Auger spectroscopy for the in situ elemental characterization of sub-micrometer presolar grains. *Meteorit. Planet. Sci.* **44**, 1033–1049.

Stroud, R.M., Chisholm, M.F., Heck, P.R., Alexander, C.M.O'D. and Nittler, L.R. (2011) Supernova shock-wave-induced co-formation of glassy carbon and nanodiamond. *Ap. J.* **738**.

Stroud, R.M., De Gregorio, B.T., Nittler, L.R. and Alexander, C.M.O'D. (2014) Comparative transmission electron microscopy studies of presolar silicate and oxide grains from the Dominion Range 08006 and Northwest Africa 5958 meteorites. Lunar Planet. Sci. XLV, #2806 (abstr.).

Stroud, R.M., Nittler, L.R. and Alexander, C.M.O'D. (2004) Polymorphism in presolar $Al_2O_3$ grains from asymptotic giant branch stars. *Science* **305**, 1455–1457.

Stroud, R.M., Nittler, L.R. and Alexander, C.M.O'D. (2013) Analytical electron microscopy of a CAI-like presolar grain and associated fine-grained matrix materials in the Dominion Range 08006 CO3 meteorite. Lunar Planet. Sci. XLIV, #2315 (abstr.).

Takigawa, A., Stroud, R.M., Nittler, L.R., Vicenzi, E.P., Herzing, A., Alexander, C.M.O'D. and Huss, G.R. (2014) Crystal structure, morphology, and isotopic compositions of presolar $Al_2O_3$ grains in unequilibrated ordinary chondrites. Lunar Planet. Sci. XLV, #1465 (abstr.).

Vollmer, C., Brenker, F.E., Hoppe, P. and Stroud, R.M. (2009a) Direct laboratory analysis of silicate stardust from red giant stars. *Ap. J.* **700**, 774–782.

Vollmer, C., Hoppe, P. and Brenker, F.E. (2013) Transmission electron microscopy of Al-rich silicate stardust from asymptotic giant branch stars. *Ap. J.* **769**, 61.

Vollmer, C., Hoppe, P., Brenker, F.E. and Holzapfel, C. (2007) Stellar $MgSiO_3$ perovskite: A shock-transformed stardust silicate found in a meteorite. *Ap. J.* **666**, L49–L52.

Vollmer, C., Hoppe, P., Stadermann, F.J., Floss, C. and Brenker, F.E. (2009b) NanoSIMS analysis and Auger electron spectroscopy of silicate and oxide stardust from the carbonaceous chondrite Acfer 094. *Geochim. Cosmochim. Acta* **73**, 7127–7149.

Vollmer, C., Kepaptsoglou, D., Leitner, J., Busemann, H., Spring, N.H., Ramasse, Q.M., Hoppe, P. and Nittler, L.R. (2014) Fluid-induced organic synthesis in the solar nebula recorded in extraterrestrial dust from meteorites. *Proc. Nat. Acad. Sci. USA* **111**, 15338–15343.

Wasserburg, G.J., Boothroyd, A.I. and Sackmann, I.-J. (1995) Deep circulation in red giant stars: A solution to the carbon and oxygen isotope puzzles? *Ap. J.* **447**, L37–40.

Yada, T., Floss, C., Stadermann, F.J., Zinner, E., Nakamura, T., Noguchi, T. and Lea, A.S. (2008) Stardust in Antarctic micrometeorites. *Meteorit. Planet. Sci.* **43**, 1287–1298.

Yang, J. and Epstein, S. (1983) Interstellar organic matter in meteorites. *Geochim. Cosmochim. Acta* **47**, 2199–2216.

Zega, T.J., Alexander, C.M.O'D., Nittler, L.R. and Stroud, R.M. (2011) A transmission electron microscopy study of presolar hibonite. *Ap. J.* **730**, 83–92.

Zega, T.J., Nittler, L.R., Gyngard, F., Alexander, C.M.O'D., Stroud, R.M. and Zinner, E.K. (2014) A transmission electron microscopy study of presolar spinel. *Geochim. Cosmochim. Acta* **124**, 152–169.

Zhao, X., Floss, C., Lin, Y. and Bose, M. (2013) Stardust investigation into the CR chondrite Grove Mountain 021710. *Ap. J.* **769**.

Zinner, E. (2014) 1.4 - Presolar grains, in: Davis, A.M. (Ed.), Meteorites and Cosmochemical Processes (Vol. 1), Treatise on Geochemistry (Second Edition, eds: H. D. Holland and K. K. Turekian). Elsevier-Pergamon, Oxford, pp. 181–213.

Zinner, E., Amari, S., Guinness, R., Nguyen, A., Stadermann, F.J., Walker, R.M. and Lewis, R.S. (2003) Presolar spinel grains from the Murray and Murchison carbonaceous chondrites. *Geochim. Cosmochim. Acta* **67**, 5083–5095.

Table 1 Areas analyzed by NanoSIMS, resultant number of presolar grains and matrix-normalized abundances of O-rich presolar grains.

| | Analyzed Area ($\mu m^2$) | # Presolar Grains | | Abundance of presolar O-rich grains (ppm) | | | | |
|---|---|---|---|---|---|---|---|---|
| | | O-rich | SiC | Average | Error – (1σ) | Error + (1σ) | Error – (2σ) | Error + (2σ) |
| Area 2 | 10,820 | 43 | 7 | 299 | | | | |
| Area 3 | 7,500 | 36 | 5 | 213 | | | | |
| Area 6 | 3,000 | 11 | 0 | 217 | | | | |
| Area 7 | 6,550 | 11 | 5 | 155 | | | | |
| Total | | 101 | 17 | 257[a] | 41 | 44 | 76 | 96 |

[a] Excluding Area 7; errors estimated from Monte Carlo calculations (see text)

Table 2. Sizes and isotopic compositions of presolar O-rich grains.

| Grain | Size (nm) | Size (nm, corrected) | Group | $^{17}O/^{16}O$ (× $10^{-4}$) | $^{18}O/^{16}O$ (× $10^{-3}$) | $Si^-/O^-$ (× $10^{-3}$) | $AlO^-/O^-$ (× $10^{-4}$) |
|---|---|---|---|---|---|---|---|
| DOM-1 | 207 | 168 | 1 | 5.30 ± 0.26 | 1.990 ± 0.049 | 8.81 ± 0.10 | 4.5 ± 0.2 |
| DOM-2 | 237 | 205 | 1 | 5.26 ± 0.24 | 1.944 ± 0.046 | 13.50 ± 0.12 | 4.4 ± 0.2 |
| DOM-3 | 545 | 532 | 1 | 9.91 ± 0.18 | 1.820 ± 0.023 | 10.21 ± 0.05 | 13.2 ± 0.2 |
| DOM-4 | 229 | 195 | 1 | 5.49 ± 0.29 | 1.892 ± 0.056 | 7.66 ± 0.11 | 10.9 ± 0.4 |
| DOM-5 | 192 | 150 | 3 | 3.42 ± 0.26 | 1.700 ± 0.057 | 8.81 ± 0.12 | 2.0 ± 0.2 |
| DOM-6 | 261 | 232 | 1 | 5.56 ± 0.28 | 2.010 ± 0.050 | 7.38 ± 0.10 | 25.6 ± 0.6 |
| DOM-7 | 275 | 248 | 1 | 6.09 ± 0.32 | 1.348 ± 0.055 | 10.14 ± 0.13 | 5.1 ± 0.3 |
| DOM-8 | 522 | 508 | 1 | 8.83 ± 0.17 | 1.972 ± 0.025 | 10.19 ± 0.06 | 23.2 ± 0.3 |
| DOM-9 | 241 | 209 | 1 | 6.34 ± 0.39 | 1.978 ± 0.066 | 9.19 ± 0.14 | 10.0 ± 0.5 |
| DOM-10 | 333 | 310 | 1 | 6.14 ± 0.30 | 1.949 ± 0.051 | 5.63 ± 0.09 | 41.8 ± 0.7 |
| DOM-11 | 938 | 930 | 1 | 4.97 ± 0.18 | 1.917 ± 0.034 | 7.74 ± 0.07 | 21.7 ± 0.4 |
| DOM-12 | 382 | 362 | 1 | 6.21 ± 0.17 | 1.576 ± 0.029 | 16.89 ± 0.09 | 13.6 ± 0.2 |
| DOM-13 | 257 | 227 | 1 | 7.05 ± 0.24 | 2.006 ± 0.039 | 13.88 ± 0.10 | 15.1 ± 0.3 |
| DOM-14 | 282 | 255 | 1 | 8.20 ± 0.31 | 1.994 ± 0.046 | 11.96 ± 0.11 | 15.0 ± 0.4 |
| DOM-15 | 268 | 240 | 1 | 5.03 ± 0.21 | 2.043 ± 0.041 | 13.05 ± 0.11 | 16.7 ± 0.4 |
| DOM-16 | 330 | 307 | 1 | 6.57 ± 0.19 | 1.473 ± 0.031 | 12.21 ± 0.08 | 28.2 ± 0.4 |
| DOM-17 | 363 | 343 | 1 | 23.63 ± 0.57 | 2.050 ± 0.051 | 17.57 ± 0.15 | 7.8 ± 0.3 |
| DOM-18 | 324 | 301 | 1 | 6.75 ± 0.21 | 1.325 ± 0.035 | 15.20 ± 0.10 | 24.6 ± 0.4 |
| DOM-19 | 225 | 190 | 1 | 5.89 ± 0.34 | 1.886 ± 0.060 | 18.11 ± 0.18 | 8.4 ± 0.4 |
| DOM-20 | 176 | 129 | 4 | 5.97 ± 0.48 | 2.259 ± 0.088 | 15.90 ± 0.24 | 7.2 ± 0.5 |
| DOM-21 | 369 | 349 | 1 | 7.47 ± 0.37 | 1.989 ± 0.057 | 15.11 ± 0.16 | 3.5 ± 0.2 |
| DOM-22 | 245 | 214 | 1 | 6.90 ± 0.33 | 1.993 ± 0.053 | 12.65 ± 0.13 | 8.4 ± 0.3 |
| DOM-23 | 268 | 240 | 2 | 5.76 ± 0.36 | 1.048 ± 0.063 | 12.27 ± 0.16 | 42.7 ± 0.9 |
| DOM-24 | 361 | 340 | 1 | 6.21 ± 0.16 | 1.944 ± 0.028 | 8.88 ± 0.06 | 120.5 ± 0.7 |
| DOM-25 | 253 | 223 | 1 | 5.54 ± 0.22 | 1.967 ± 0.039 | 10.62 ± 0.09 | 180.7 ± 1.2 |
| DOM-26 | 257 | 227 | 1 | 5.47 ± 0.42 | 1.604 ± 0.077 | 15.39 ± 0.22 | 15.9 ± 0.7 |
| DOM-27 | 245 | 214 | 1 | 7.17 ± 0.28 | 2.092 ± 0.045 | 11.93 ± 0.11 | 37.4 ± 0.6 |
| DOM-28 | 245 | 214 | 1 | 5.30 ± 0.23 | 1.891 ± 0.043 | 13.33 ± 0.11 | 73.1 ± 0.8 |
| DOM-29 | 350 | 329 | 4 | 3.72 ± 0.19 | 2.490 ± 0.047 | 9.33 ± 0.09 | 5.2 ± 0.2 |
| DOM-30 | 176 | 129 | 1 | 3.78 ± 0.46 | 2.573 ± 0.123 | 5.11 ± 0.18 | 27.0 ± 1.3 |
| DOM-31 | 384 | 365 | 1 | 8.12 ± 0.20 | 1.866 ± 0.032 | 10.57 ± 0.07 | 15.3 ± 0.3 |
| DOM-32 | 318 | 294 | 1 | 8.60 ± 0.25 | 2.001 ± 0.039 | 8.06 ± 0.08 | 22.3 ± 0.4 |
| DOM-33 | 427 | 410 | 1 | 9.11 ± 0.24 | 1.724 ± 0.037 | 4.60 ± 0.06 | 108.6 ± 0.9 |
| DOM-34 | 268 | 240 | 1 | 5.37 ± 0.23 | 1.636 ± 0.046 | 7.32 ± 0.09 | 12.5 ± 0.4 |
| DOM-35 | 257 | 227 | 1 | 4.92 ± 0.22 | 2.066 ± 0.046 | 4.84 ± 0.07 | 46.1 ± 0.7 |
| DOM-36 | 275 | 248 | 1 | 6.29 ± 0.27 | 1.918 ± 0.050 | 8.55 ± 0.11 | 24.3 ± 0.6 |
| DOM-37 | 330 | 307 | 1 | 7.08 ± 0.28 | 1.966 ± 0.048 | 9.59 ± 0.11 | 6.4 ± 0.3 |
| DOM-38 | 275 | 248 | 1 | 5.21 ± 0.22 | 2.013 ± 0.044 | 7.37 ± 0.09 | 34.5 ± 0.6 |
| DOM-39 | 279 | 252 | 1 | 6.97 ± 0.27 | 1.868 ± 0.046 | 8.06 ± 0.10 | 31.2 ± 0.6 |
| DOM-40 | 268 | 240 | 1 | 5.07 ± 0.28 | 1.512 ± 0.057 | 9.45 ± 0.13 | 9.8 ± 0.4 |
| DOM-41 | 305 | 281 | 1 | 6.25 ± 0.21 | 2.034 ± 0.038 | 10.97 ± 0.10 | 57.1 ± 0.7 |
| DOM-42 | 245 | 214 | 4 | 5.71 ± 0.33 | 1.610 ± 0.064 | 9.92 ± 0.16 | 11.8 ± 0.5 |
| DOM-43 | 253 | 223 | 4 | 3.76 ± 0.23 | 2.365 ± 0.058 | 7.78 ± 0.11 | 17.6 ± 0.5 |
| DOM-44 | 241 | 209 | 1 | 6.43 ± 0.42 | 1.635 ± 0.073 | 12.14 ± 0.18 | *n.m.* |
| DOM-45 | 282 | 255 | 1 | 6.18 ± 0.18 | 1.654 ± 0.031 | 8.84 ± 0.06 | *n.m.* |
| DOM-46 | 229 | 195 | 4 | 5.22 ± 0.24 | 2.685 ± 0.053 | 16.53 ± 0.12 | *n.m.* |
| DOM-47 | 257 | 227 | 1 | 6.16 ± 0.19 | 1.576 ± 0.034 | 8.76 ± 0.07 | *n.m.* |
| DOM-48 | 436 | 420 | 1 | 4.70 ± 0.11 | 1.973 ± 0.022 | 12.80 ± 0.06 | *n.m.* |
| DOM-49 | 220 | 185 | 1 | 5.10 ± 0.30 | 2.052 ± 0.059 | 19.24 ± 0.15 | *n.m.* |
| DOM-50 | 253 | 223 | 1 | 5.91 ± 0.25 | 2.009 ± 0.045 | 12.67 ± 0.10 | *n.m.* |
| DOM-51 | 296 | 270 | 1 | 4.44 ± 0.22 | 1.657 ± 0.045 | 22.83 ± 0.12 | *n.m.* |
| DOM-52 | 299 | 274 | 1 | 6.02 ± 0.19 | 1.962 ± 0.034 | 17.33 ± 0.09 | *n.m.* |
| DOM-53 | 220 | 185 | 1 | 5.13 ± 0.19 | 1.946 ± 0.037 | 5.58 ± 0.06 | *n.m.* |
| DOM-54 | 296 | 270 | 1 | 5.40 ± 0.19 | 1.926 ± 0.036 | 17.33 ± 0.10 | *n.m.* |
| DOM-55 | 197 | 156 | 4 | 4.36 ± 0.40 | 2.449 ± 0.095 | 14.02 ± 0.23 | *n.m.* |
| DOM-56 | 176 | 129 | 1 | 6.22 ± 0.45 | 2.017 ± 0.080 | 7.12 ± 0.15 | *n.m.* |
| DOM-57 | 207 | 168 | 4 | 4.30 ± 0.34 | 2.488 ± 0.082 | 7.51 ± 0.14 | *n.m.* |
| DOM-58 | 237 | 205 | 4 | 4.61 ± 0.57 | 3.101 ± 0.146 | 11.88 ± 0.29 | *n.m.* |
| DOM-59 | 225 | 190 | 1 | 5.34 ± 0.29 | 1.964 ± 0.055 | 2.56 ± 0.06 | *n.m.* |
| DOM-60 | 202 | 162 | 4 | 3.18 ± 0.44 | 2.809 ± 0.119 | 1.47 ± 0.09 | *n.m.* |
| DOM-61 | 344 | 323 | 1 | 4.73 ± 0.12 | 2.042 ± 0.024 | 12.32 ± 0.06 | *n.m.* |
| DOM-62 | 253 | 223 | 1 | 5.45 ± 0.19 | 1.684 ± 0.036 | 10.53 ± 0.08 | *n.m.* |
| DOM-63 | 344 | 323 | 1 | 7.58 ± 0.21 | 1.964 ± 0.033 | 13.22 ± 0.09 | *n.m.* |
| DOM-64 | 153 | 094 | 4 | 3.35 ± 0.44 | 2.559 ± 0.111 | 12.59 ± 0.25 | *n.m.* |
| DOM-65 | 220 | 185 | 4 | 4.05 ± 0.24 | 2.664 ± 0.059 | 14.01 ± 0.14 | *n.m.* |
| DOM-66 | 268 | 240 | 1 | 8.21 ± 0.26 | 1.909 ± 0.039 | 8.16 ± 0.08 | *n.m.* |
| DOM-67 | 153 | 094 | 4 | 4.12 ± 0.26 | 2.248 ± 0.059 | 13.93 ± 0.15 | *n.m.* |
| DOM-68 | 292 | 267 | 1 | 5.94 ± 0.17 | 1.628 ± 0.031 | 16.49 ± 0.10 | *n.m.* |
| DOM-69 | 197 | 156 | 1 | 7.70 ± 0.42 | 1.974 ± 0.066 | 20.68 ± 0.23 | *n.m.* |
| DOM-70 | 299 | 274 | 1 | 4.75 ± 0.20 | 1.671 ± 0.039 | 17.33 ± 0.12 | *n.m.* |
| DOM-71 | 272 | 244 | 1 | 6.08 ± 0.18 | 1.910 ± 0.032 | 7.22 ± 0.06 | *n.m.* |
| DOM-72 | 197 | 156 | 4 | 3.69 ± 0.21 | 2.285 ± 0.051 | 14.38 ± 0.14 | *n.m.* |
| DOM-73 | 253 | 223 | 1 | 6.39 ± 0.34 | 2.096 ± 0.059 | 20.37 ± 0.20 | *n.m.* |
| DOM-74 | 257 | 227 | 1 | 5.61 ± 0.20 | 1.973 ± 0.037 | 11.20 ± 0.09 | *n.m.* |
| DOM-75 | 220 | 185 | 4 | 3.86 ± 0.18 | 2.709 ± 0.047 | 10.52 ± 0.09 | *n.m.* |
| DOM-76 | 253 | 223 | 1 | 6.24 ± 0.22 | 1.481 ± 0.038 | 13.91 ± 0.10 | *n.m.* |
| DOM-77 | 369 | 349 | 1 | 6.01 ± 0.16 | 1.259 ± 0.052 | 15.10 ± 0.08 | *n.m.* |
| DOM-78 | 264 | 236 | 1 | 6.93 ± 0.32 | 1.915 ± 0.060 | 13.57 ± 0.11 | *n.m.* |
| DOM-79 | 416 | 398 | 1 | 5.41 ± 0.13 | 2.067 ± 0.042 | 12.25 ± 0.06 | *n.m.* |
| DOM-80 | 416 | 398 | 1 | 8.17 ± 0.16 | 2.056 ± 0.026 | 5.02 ± 0.04 | 32.6 ± 0.3 |
| DOM-81 | 279 | 252 | 1 | 6.40 ± 0.20 | 1.460 ± 0.037 | 6.25 ± 0.06 | 26.1 ± 0.4 |
| DOM-82 | 211 | 174 | 1 | 5.93 ± 0.26 | 2.053 ± 0.050 | 5.02 ± 0.08 | 79.7 ± 1.0 |
| DOM-83 | 341 | 320 | 1 | 8.98 ± 0.20 | 1.810 ± 0.030 | 5.95 ± 0.05 | 37.8 ± 0.4 |
| DOM-84 | 237 | 205 | 1 | 5.37 ± 0.25 | 1.925 ± 0.050 | 5.96 ± 0.09 | 43.0 ± 0.7 |
| DOM-85 | 299 | 274 | 1 | 15.41 ± 0.83 | 2.082 ± 0.099 | 7.00 ± 0.18 | 51.4 ± 1.5 |
| DOM-86 | 182 | 136 | 1 | 4.72 ± 0.21 | 2.173 ± 0.046 | 2.79 ± 0.05 | 77.6 ± 0.9 |
| DOM-87 | 302 | 277 | 1 | 5.09 ± 0.15 | 1.951 ± 0.031 | 2.82 ± 0.04 | 35.6 ± 0.4 |
| DOM-88 | 253 | 223 | 1 | 5.13 ± 0.18 | 1.957 ± 0.037 | 3.78 ± 0.05 | 28.3 ± 0.4 |
| DOM-89 | 402 | 383 | 2 | 7.80 ± 0.18 | 1.124 ± 0.031 | 4.46 ± 0.05 | 223.5 ± 1.0 |
| DOM-90 | 279 | 252 | 1 | 5.65 ± 0.20 | 1.935 ± 0.039 | 4.99 ± 0.06 | 27.5 ± 0.5 |
| DOM-91 | 296 | 270 | 1 | 5.54 ± 0.20 | 1.725 ± 0.038 | 3.60 ± 0.05 | 49.4 ± 0.6 |
| DOM-92 | 220 | 185 | 1 | 4.58 ± 0.26 | 1.655 ± 0.056 | 2.82 ± 0.07 | 93.7 ± 1.2 |
| DOM-93 | 207 | 168 | 1 | 5.37 ± 0.29 | 2.105 ± 0.057 | 4.79 ± 0.09 | 73.6 ± 1.1 |
| DOM-94 | 358 | 337 | 1 | 5.25 ± 0.14 | 1.917 ± 0.028 | 2.93 ± 0.04 | 83.1 ± 0.6 |
| DOM-95 | 192 | 150 | 4 | 3.44 ± 0.31 | 2.368 ± 0.077 | 4.07 ± 0.11 | 48.6 ± 1.2 |
| DOM-96 | 533 | 519 | 4 | 3.51 ± 0.17 | 4.773 ± 0.060 | 3.30 ± 0.05 | 2.2 ± 0.1 |
| DOM-97 | 341 | 320 | 1 | 5.14 ± 0.18 | 1.952 ± 0.035 | 4.16 ± 0.05 | 158.9 ± 1.0 |
| DOM-98 | 387 | 368 | 1 | 5.57 ± 0.17 | 1.477 ± 0.032 | 5.68 ± 0.06 | 139.7 ± 0.9 |
| DOM-99 | 523 | 509 | 1 | 6.03 ± 0.14 | 1.891 ± 0.025 | 4.03 ± 0.04 | 38.5 ± 0.4 |
| DOM-100 | 377 | 357 | 1 | 6.61 ± 0.19 | 1.372 ± 0.034 | 4.21 ± 0.05 | 43.1 ± 0.5 |
| DOM-101 | 379 | 360 | 1 | 6.17 ± 0.22 | 1.619 ± 0.039 | 7.23 ± 0.08 | 23.2 ± 0.5 |

Table 3 Sizes and isotopic compositions of C-anomalous grains.

| Grain | Phase | Diameter (nm) | $\delta^{13}C$ | $\delta^{15}N$ |
|---|---|---|---|---|
| A3b2_4 | SiC | 117 | 2020 ± 156 | n. m. |
| A3_9 | SiC | 205 | 831 ± 46 | n. m. |
| A3b2_4b | SiC | 94 | 336 ± 76 | n. m. |
| A3_11 | SiC | 263 | 304 ± 29 | n. m. |
| A2_6 | SiC | 214 | 305 ± 49 | n. m. |
| A2d_3 | SiC | 263 | 1026 ± 51 | n. m. |
| A2r2_20 | SiC | 209 | 493 ± 109 | n. m. |
| A2r_3 | SiC | 168 | 514 ± 110 | n. m. |
| A2r_16 | SiC | 129 | 1034 ± 209 | n. m. |
| A2r2_15 | SiC | 121 | 515 ± 142 | n. m. |
| A2r_17 | SiC | 368 | 368 ± 60 | n. m. |
| A7_16 | SiC | 354 | 276 ± 27 | n. m. |
| A7_49 | SiC | 252 | 348 ± 58 | n. m. |
| A7_59 | SiC | 388 | 445 ± 48 | n. m. |
| A7_66 | SiC | 420 | 208 ± 30 | n. m. |
| A7_41 | SiC | 413 | 2240 ± 90 | n. m. |
| A3cCN_7 | SiC | 252 | 251 ± 42 | -73 ± 163 |
| A2c_20 | organic | 195 | -294 ± 45 | n. m. |
| A2r_3 | organic | 113 | -493 ± 111 | n. m. |
| A2r2_15 | organic | 291 | -208 ± 23 | n. m. |
| A3_14a | organic | 277 | 130 ± 17 | n. m. |
| A3_14b | organic | 185 | -203 ± 30 | n. m. |
| A3_20 | organic | 232 | -185 ± 24 | n. m. |
| A3_22 | organic | 236 | 143 ± 24 | n. m. |
| A3b2_10 | organic | 190 | -163 ± 26 | n. m. |
| A3b2_11 | organic | 252 | -135 ± 22 | n. m. |
| A3b2_16 | organic | 317 | -237 ± 25 | n. m. |
| A3b2_17 | organic | 263 | 130 ± 22 | n. m. |
| A3b2_29 | organic | 259 | 137 ± 20 | n. m. |
| A3b2_30 | organic | 304 | 200 ± 20 | n. m. |
| A3b2_5 | organic | 294 | 128 ± 17 | n. m. |
| A3cCN_1a | organic | 236 | 3 ± 17 | -130 ± 42 |
| A3cCN_1b | organic | 370 | 51 ± 17 | 268 ± 58 |
| A3cCN_2a | organic | 263 | -129 ± 38 | -115 ± 124 |
| A3cCN_2b | organic | 291 | 18 ± 15 | 529 ± 74 |
| A3cCN_2c | organic | 334 | 17 ± 15 | 213 ± 64 |
| A3cCN_2d | organic | 403 | 28 ± 21 | 298 ± 73 |
| A3cCN_2e | organic | 270 | -35 ± 38 | -294 ± 80 |
| A3cCN_3a | organic | 252 | 44 ± 16 | 670 ± 67 |
| A3cCN_3b | organic | 281 | 32 ± 13 | 496 ± 85 |
| A3cCN_3c | organic | 240 | -14 ± 25 | -248 ± 80 |
| A3cCN_3d | organic | 252 | 17 ± 27 | 340 ± 93 |
| A3cCN_3e | organic | 291 | -19 ± 29 | -210 ± 65 |
| A3cCN_3f | organic | 291 | -54 ± 36 | 449 ± 107 |
| A3cCN_4a | organic | 310 | -37 ± 14 | 659 ± 77 |
| A3cCN_4b | organic | 310 | -7 ± 20 | -188 ± 45 |
| A3cCN_4c | organic | 248 | 9 ± 24 | 871 ± 143 |
| A3cCN_4d | organic | 263 | -27 ± 24 | 338 ± 93 |
| A3cCN_5a | organic | 301 | -156 ± 46 | 184 ± 207 |
| A3cCN_5b | organic | 190 | 10 ± 21 | 459 ± 77 |
| A3cCN_5c | organic | 252 | -12 ± 19 | -256 ± 56 |
| A3cCN_5d | organic | 259 | 9 ± 41 | -297 ± 93 |
| A3cCN_6a | organic | 298 | -232 ± 13 | -225 ± 45 |
| A3cCN_6b | organic | 255 | 42 ± 16 | 342 ± 83 |
| A3cCN_6c | organic | 368 | -196 ± 14 | -215 ± 49 |
| A3cCN_6d | organic | 291 | 2 ± 16 | -101 ± 32 |
| A3cCN_6e | organic | 307 | -197 ± 17 | -183 ± 55 |
| A3cCN_6f | organic | 301 | -29 ± 29 | -234 ± 73 |
| A3cCN_6g | organic | 223 | 134 ± 24 | 543 ± 69 |
| A3cCN_6h | organic | 415 | 30 ± 33 | 337 ± 108 |
| A3cCN_6i | organic | 136 | -154 ± 45 | -47 ± 164 |
| A3cCN_6j | organic | 255 | -146 ± 41 | 273 ± 146 |
| A3cCN_7a | nanoglobule | 360 | 98 ± 9 | 361 ± 72 |
| A3cCN_7b | organic | 343 | -8 ± 10 | 288 ± 66 |
| A6b_5 | organic | 340 | -290 ± 21 | n. m. |
| A7_60 | organic | 502 | -300 ± 16 | -264 ± 29 |

Where n. m. = not measured.

Table 4: Composition and structure of presolar grains analyzed by TEM

| Grain | Composition | Structure |
|---|---|---|
| DOM-3 | $Mg_{0.45}Fe_{0.74}Ni_{0.07}Ca_{0.03}Al_{0.05}Si_{0.82}O_3$ | Amorphous |
| DOM-8 | $Mg_{1.61}Fe_{0.33}Ni_{0.03}Al_{0.02}Si_{1.00}O_4$ | Unknown (olivine?) |
| DOM-17 | $Mg_{0.33}Fe_{0.86}Ca_{0.03}Al_{0.06}Si_{0.84}O_3$ | Amorphous |
| DOM-18 | $Mg_{1.85}Fe_{0.14}Ni_{0.01}Si_{1.00}O_4$ | Olivine |
| DOM-77 | Whole grain average $Mg_{1.24}Fe_{0.43}Ni_{0.09}Al_{0.02}Ca_{0.01}Cr_{0.01}Si_{1.03}O_4$ | Concentrically zoned, polycrystalline olivine + Ca-rich nanocrystal + Al-rich oxide |

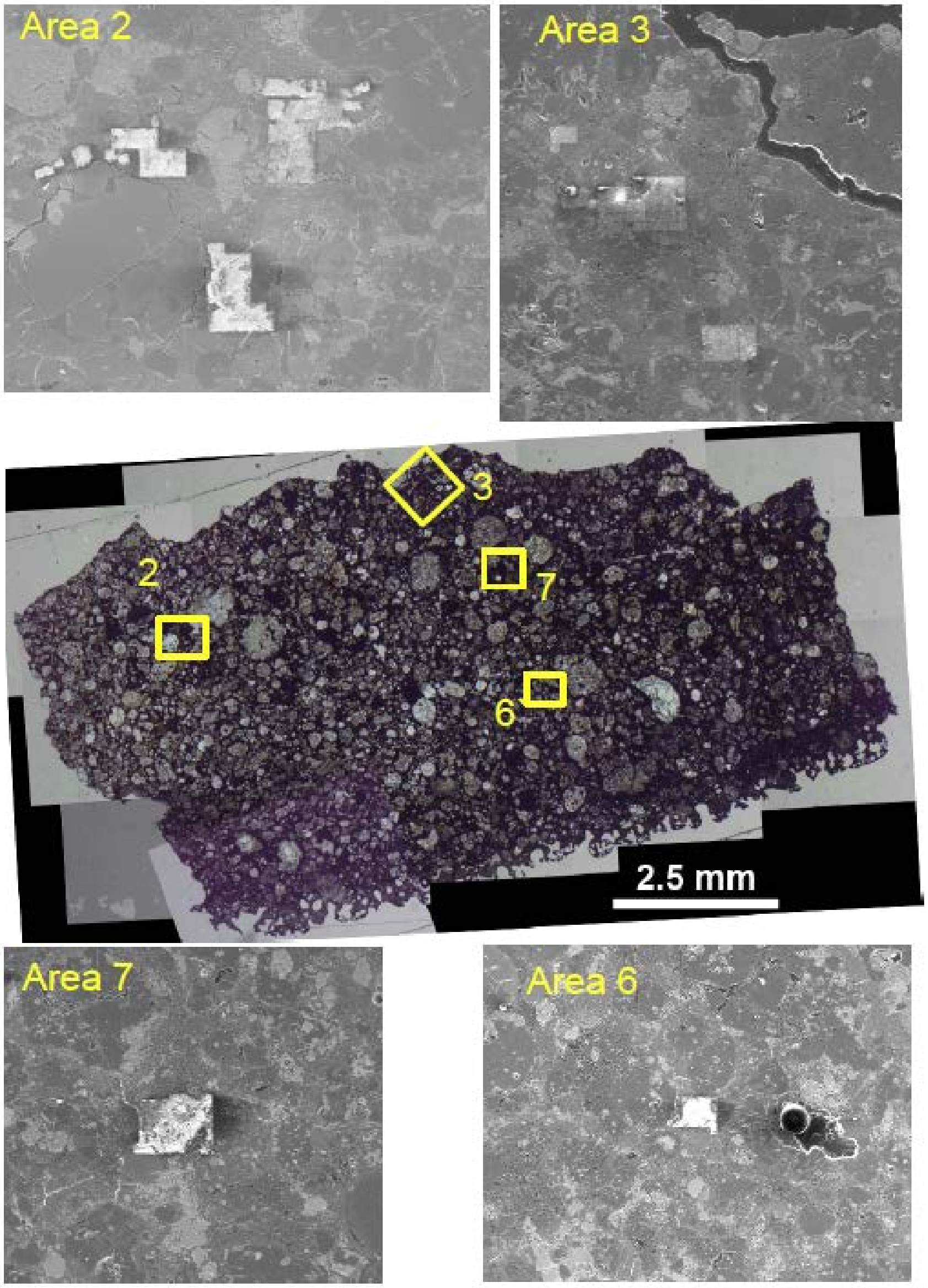

Figure 1. Reflected-light (center) and secondary electron (SE, top and bottom) images of DOM 08006 thin section. The yellow boxes and SE images indicate areas analyzed by NanoSIMS (Table 1). NanoSIMS raster areas are clearly visible in the SE images.

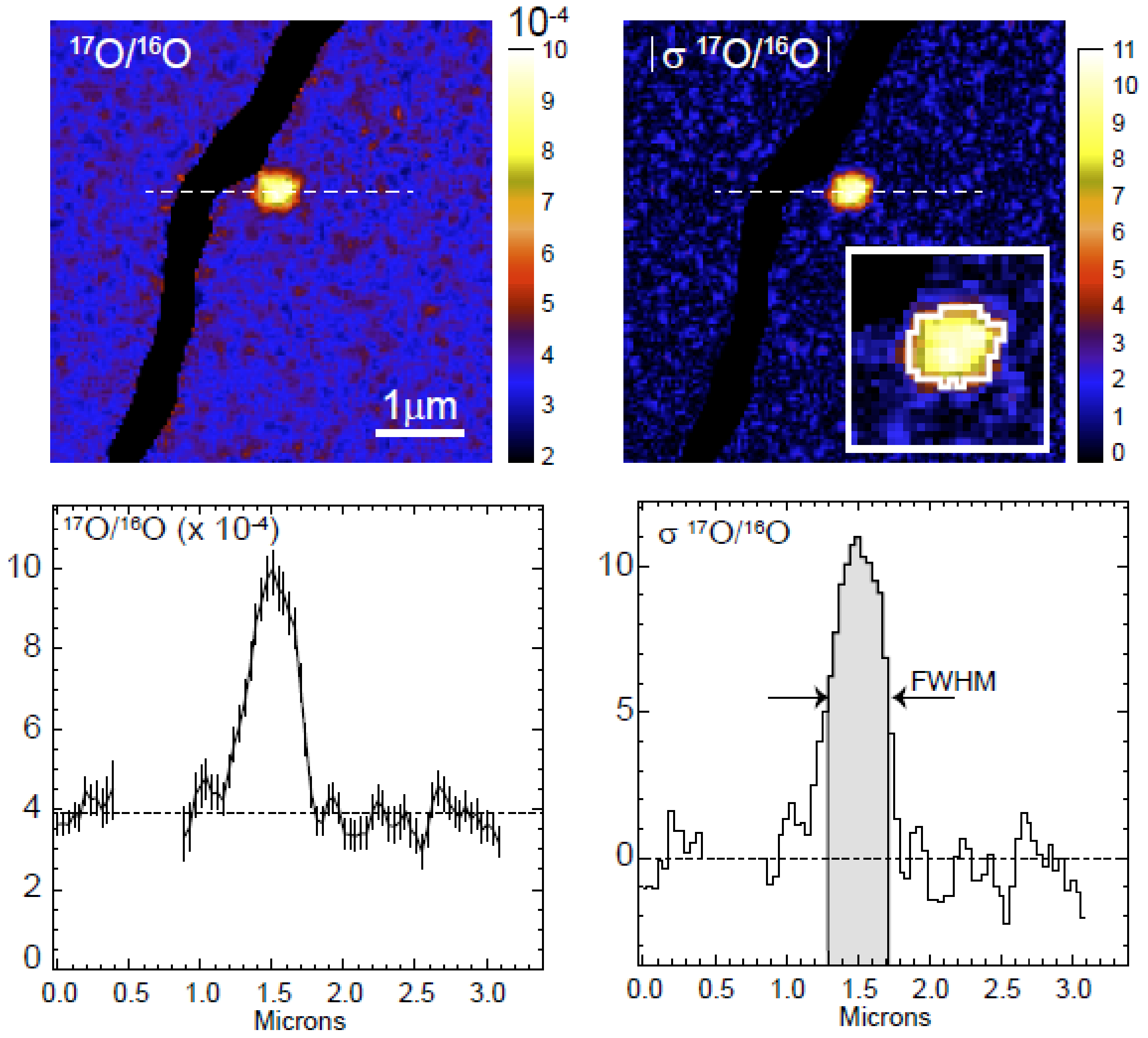

Figure 2. Left: $^{17}O/^{16}O$ ratio image and line profile data for pixels indicated by dashed line for presolar grain DOM-80. Data are smoothed with a 3-pixel boxcar, such that the ratio and error bar for a given pixel are calculated from the total counts of that pixel and its two surrounding ones. Right: corresponding "sigma" image (color scale is absolute value) and line profile, where value in each pixel is difference (in standard deviations) of measured $^{17}O/^{16}O$ ratio relative to terrestrial composition. Only pixels with values that lie within the full width at half maximum (FWHM) of the isotopically anomalous region are included in definition of the presolar grain ROI (grey area in profile plot and outline in inset).

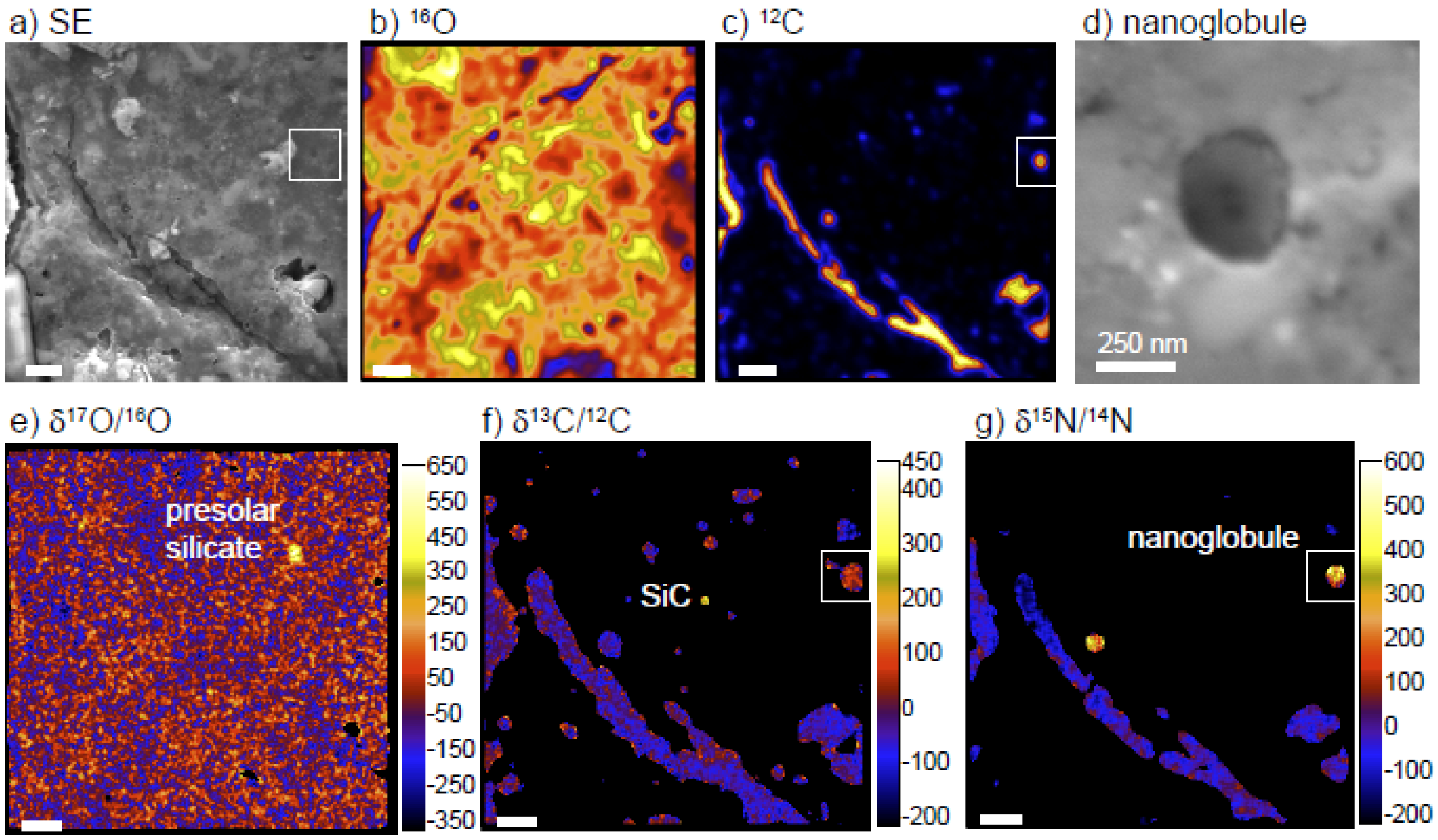

Figure 3. Example SEM and NanoSIMS images of one 10 × 10 μm area of DOM 08006. (a) SEM secondary electron image; (b) NanoSIMS $^{16}O$ ion image; (c) NanoSIMS $^{12}C$ image, (d) SEM secondary electron image of $^{13}C$- and $^{15}N$-rich organic nanoglobule, area marked by box in (a,f,g); (e) NanoSIMS $\delta^{17}O/^{16}O$ image with a $^{17}O$-rich presolar silicate (DOM-79, Table 2) indicated; (f) NanoSIMS $\delta^{12}C/^{13}C$ image with a $^{13}C$-rich presolar SiC indicated; and (g) NanoSIMS $\delta^{15}N/^{14}N$ image with two clear "hotspots" of enhanced $^{15}N$ abundance, one of which corresponds to the nanoglobule in panel (d). Unless otherwise noted, scale bars are 1 μm. Delta values are deviations from a standard ratio in permil, defined as: $\delta R = (R_{measured}/R_{standard} - 1) \times 10^3$.

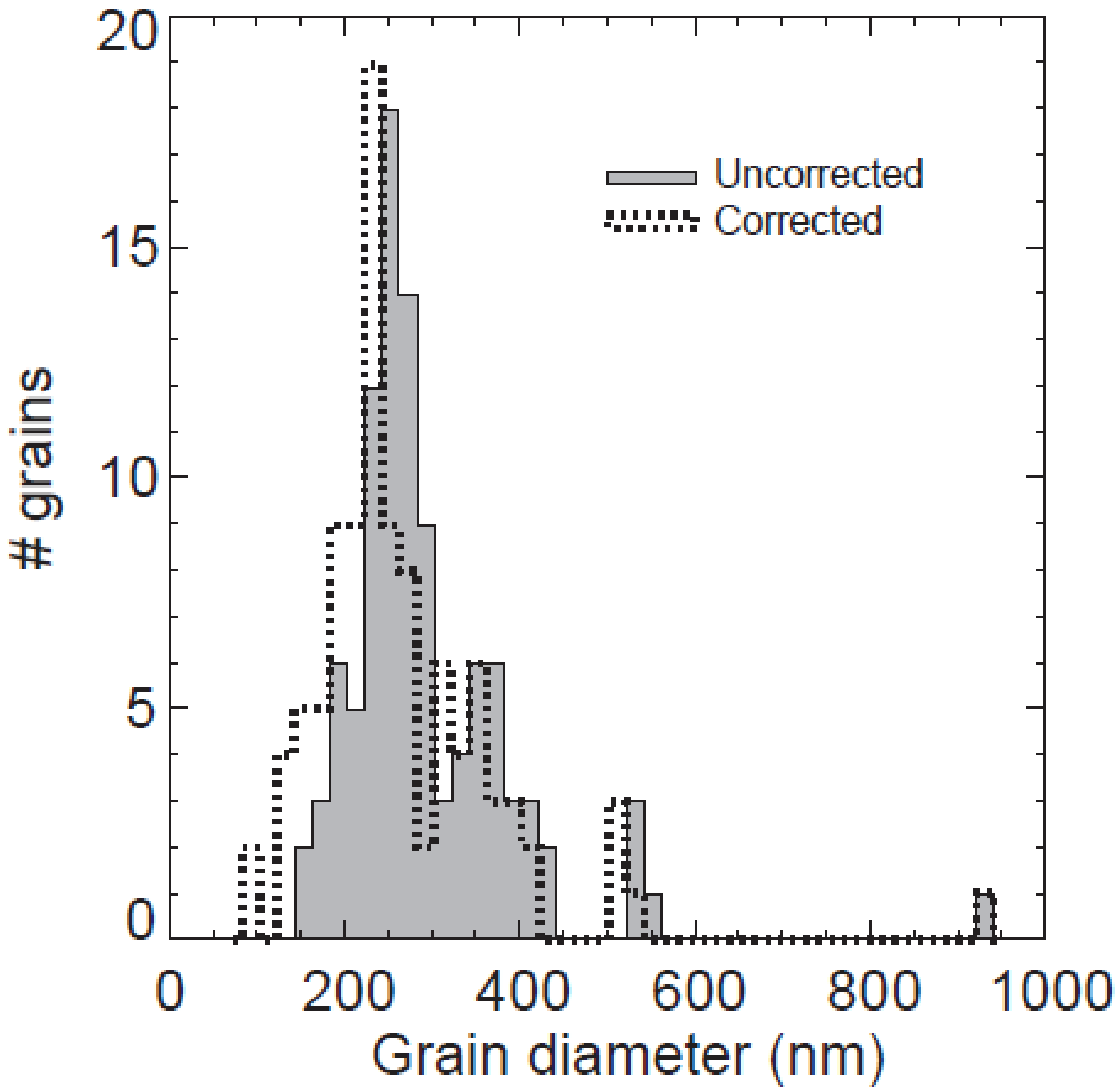

Figure 4. Size distribution of 101 presolar O-rich grains identified in DOM 08006. Grey histogram indicates measured equivalent diameters while the dotted histogram indicates data corrected for 120-nm NanoSIMS beam broadening (see text for details).

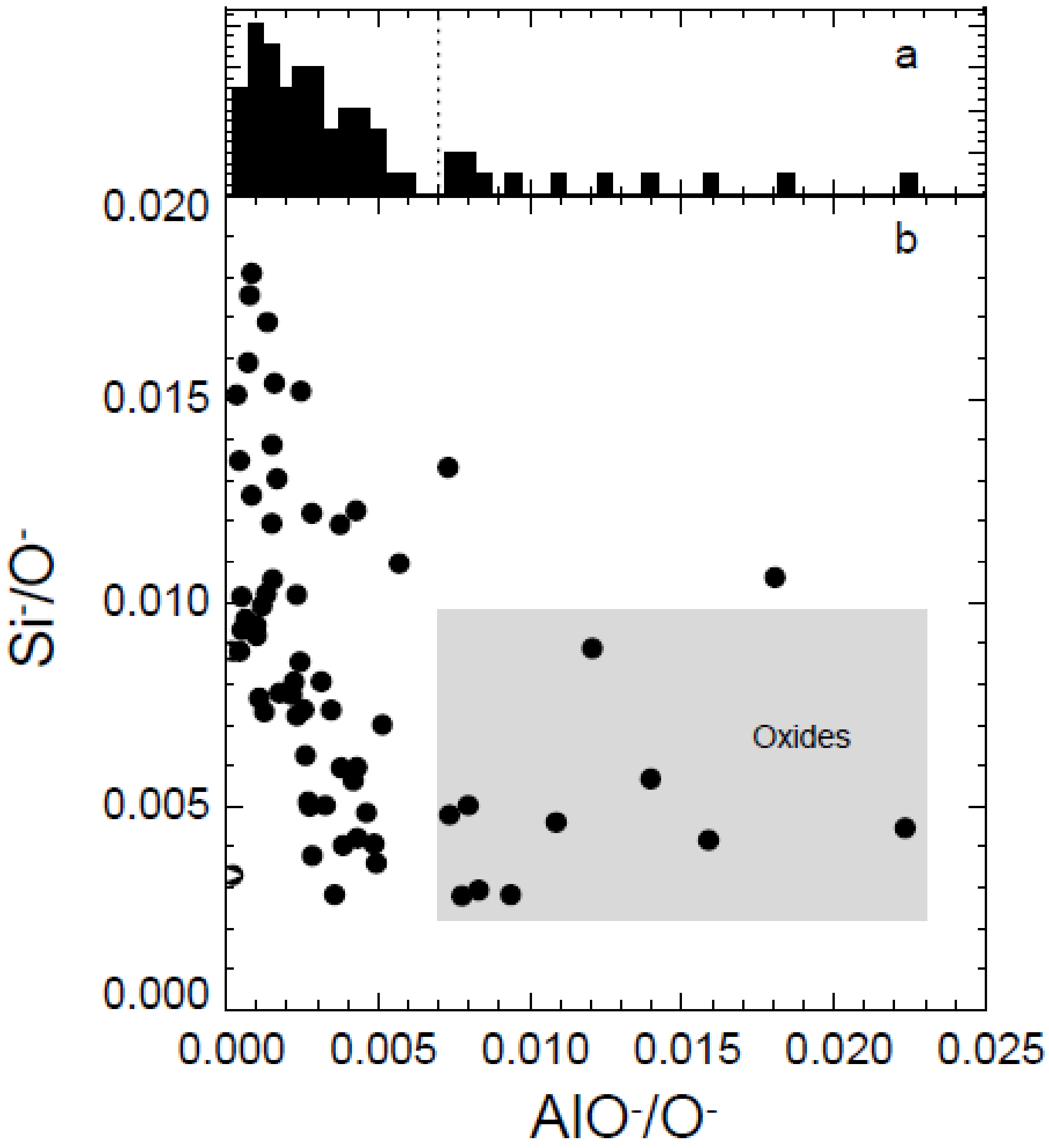

Figure 5. a) Histogram of NanoSIMS $^{27}Al^{16}O^-/^{16}O^-$ ion ratios for 65 presolar grains from DOM 08006; the maximum bin contains 8 grains. The dashed line indicates the threshold assumed to differentiate between silicate and oxide grains. b) NanoSIMS $^{28}Si^-/^{16}O^-$ secondary ion ratios plotted against $AlO^-/O^-$ ion ratios for the same presolar grains. The grey box indicates grains that are assumed to be Al-rich oxide grains, while grains with lower $AlO^-/O^-$ ratios are assumed to be silicates.

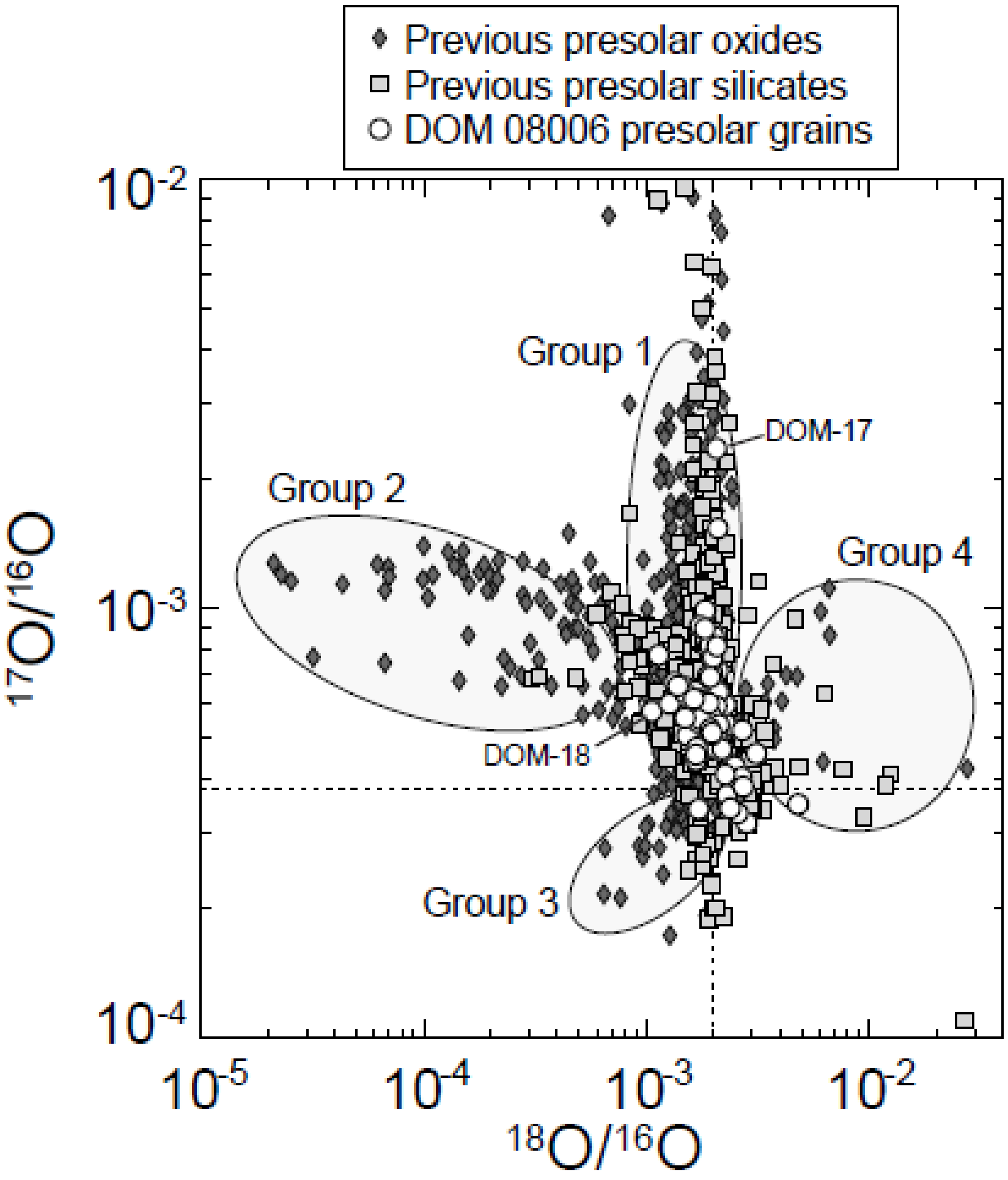

Figure 6. O-isotopic ratios in presolar O-rich grains from DOM 08006 compared to those previously reported for presolar oxides and silicates (e.g., Nittler et al., 2008; Zinner, 2014; Floss and Haenecour, 2016b). Ellipses indicate grain groups defined by Nittler et al. (1997).

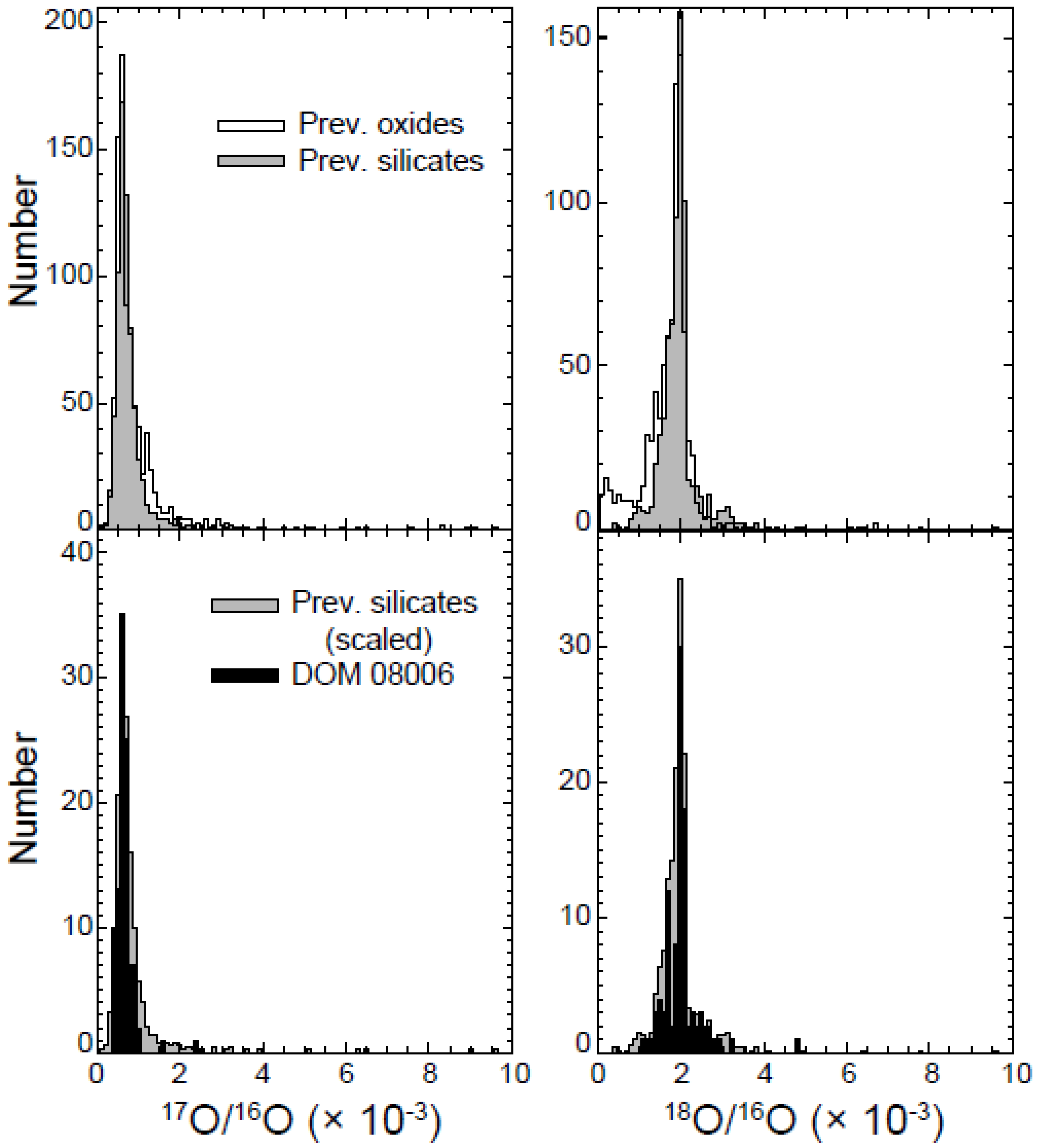

Figure 7. Histograms of O isotopic ratios measured in presolar oxide and silicate grains (see Figure 6 for data sources). Top: Previously reported data for silicates and oxides are compared; presolar silicates show a narrower distribution of $^{17}O/^{16}O$ and $^{18}O/^{16}O$ ratios than presolar oxide grains, reflecting isotope dilution in *in situ* SIMS analyses. Bottom: Data for 101 DOM 08006 presolar grains compared to the previous silicate data (re-scaled to fit on plots). The DOM 08006 data span narrower ranges of O-isotope ratios, most likely reflecting both the limited statistics as well as a conservative choice for identifying grains in ion images (see text).

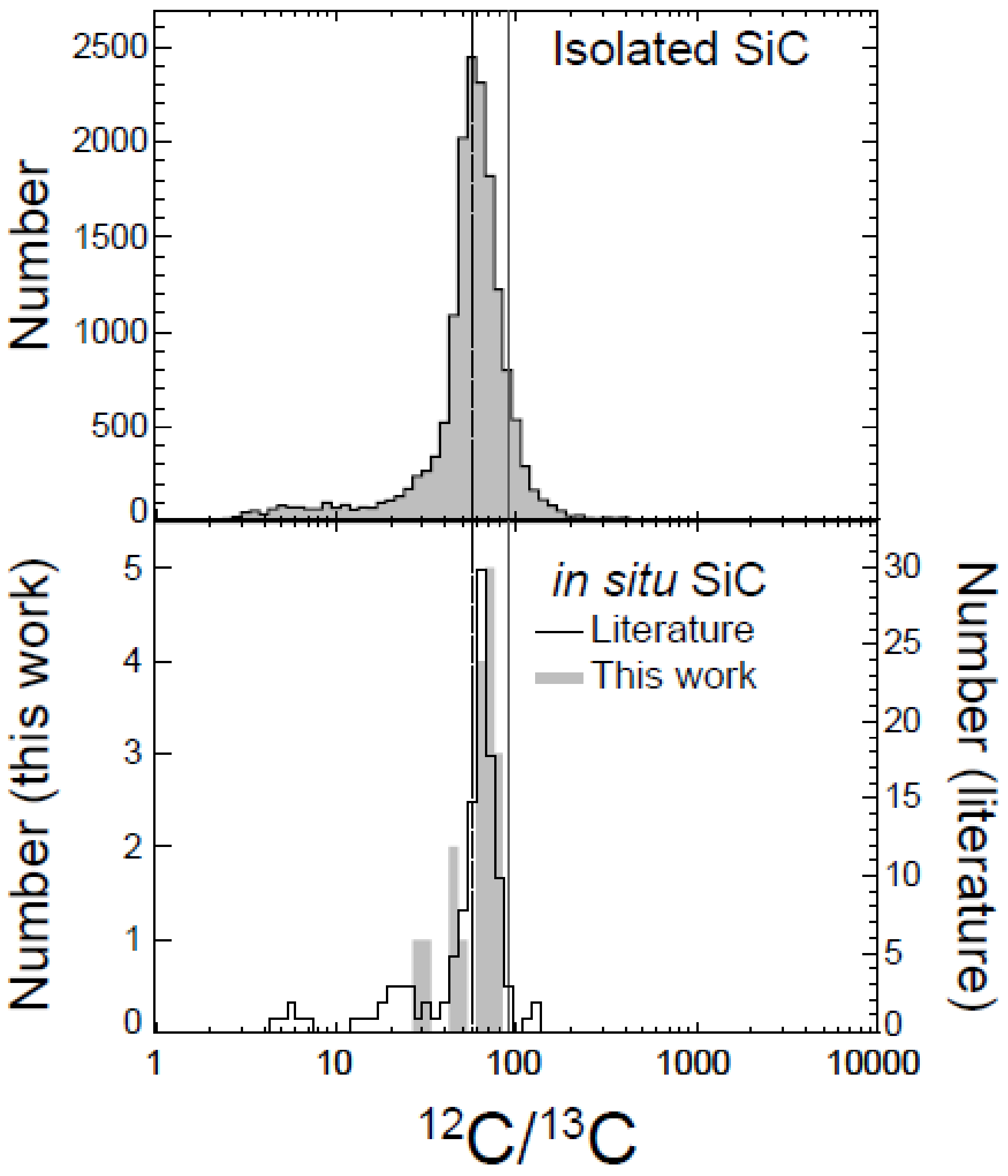

Figure 8. C-isotopic ratios of DOM 08006 presolar SiC grains compared to literature data for ~16,000 isolated grains from acid residues (top panel; Zinner, 2014) and 118 grains found in situ in previous studies (bottom panel; Floss and Stadermann, 2009a; Nguyen et al., 2010; Bose et al., 2012; Leitner et al., 2012b; Haenecour et al., 2018). The vertical solid line indicates the terrestrial ratio of 89 and the vertical dashed line indicates the peak of the isolated grain distribution at $^{12}C/^{13}C$=57.

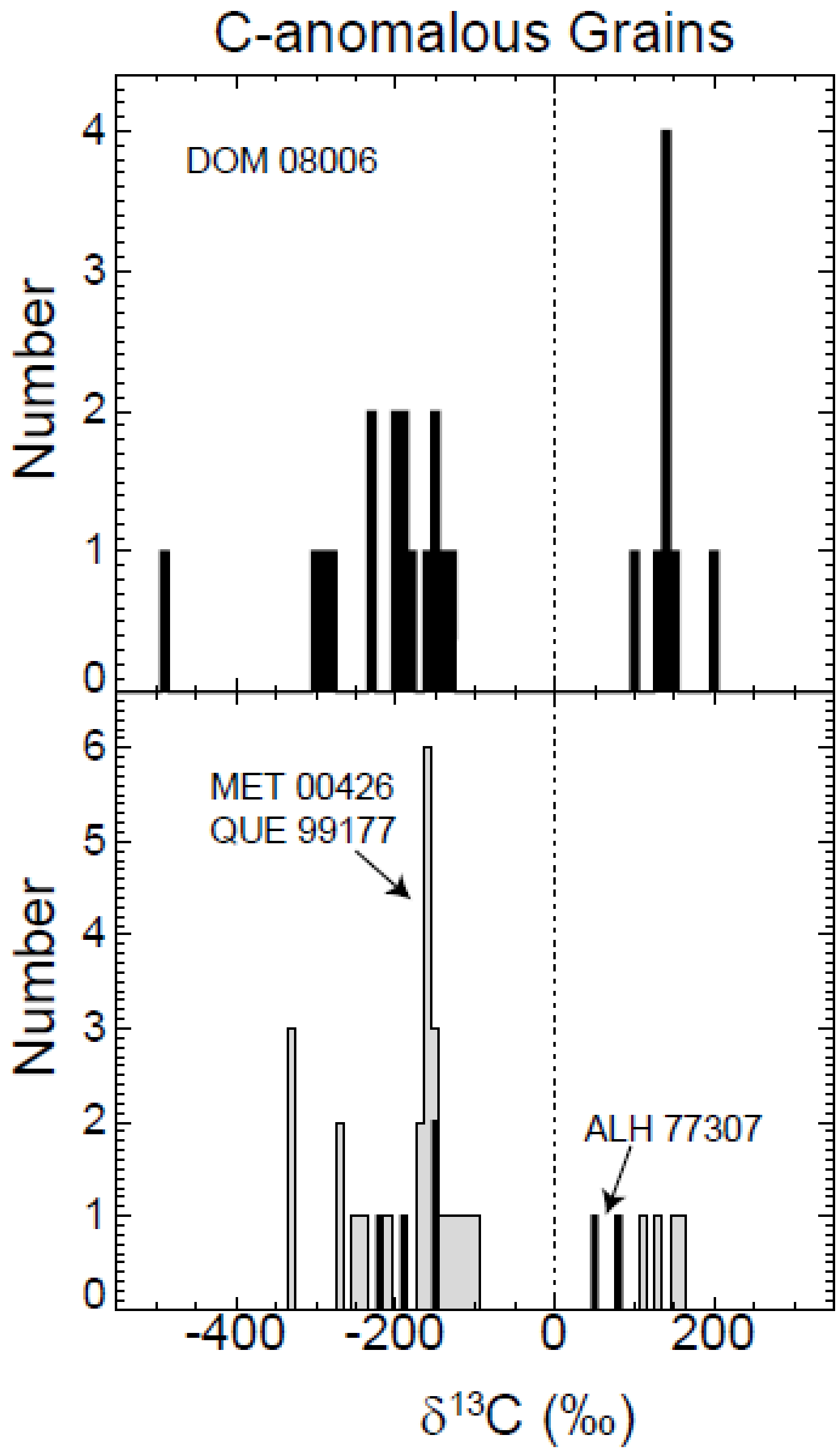

Figure 9. The distribution of $\delta^{13}$C values in 25 C-anomalous grains in DOM 00086 (top) is compared to those (bottom) measured in CR chondrites MET 00426 and QUE 99177 (Floss and Stadermann, 2009b).

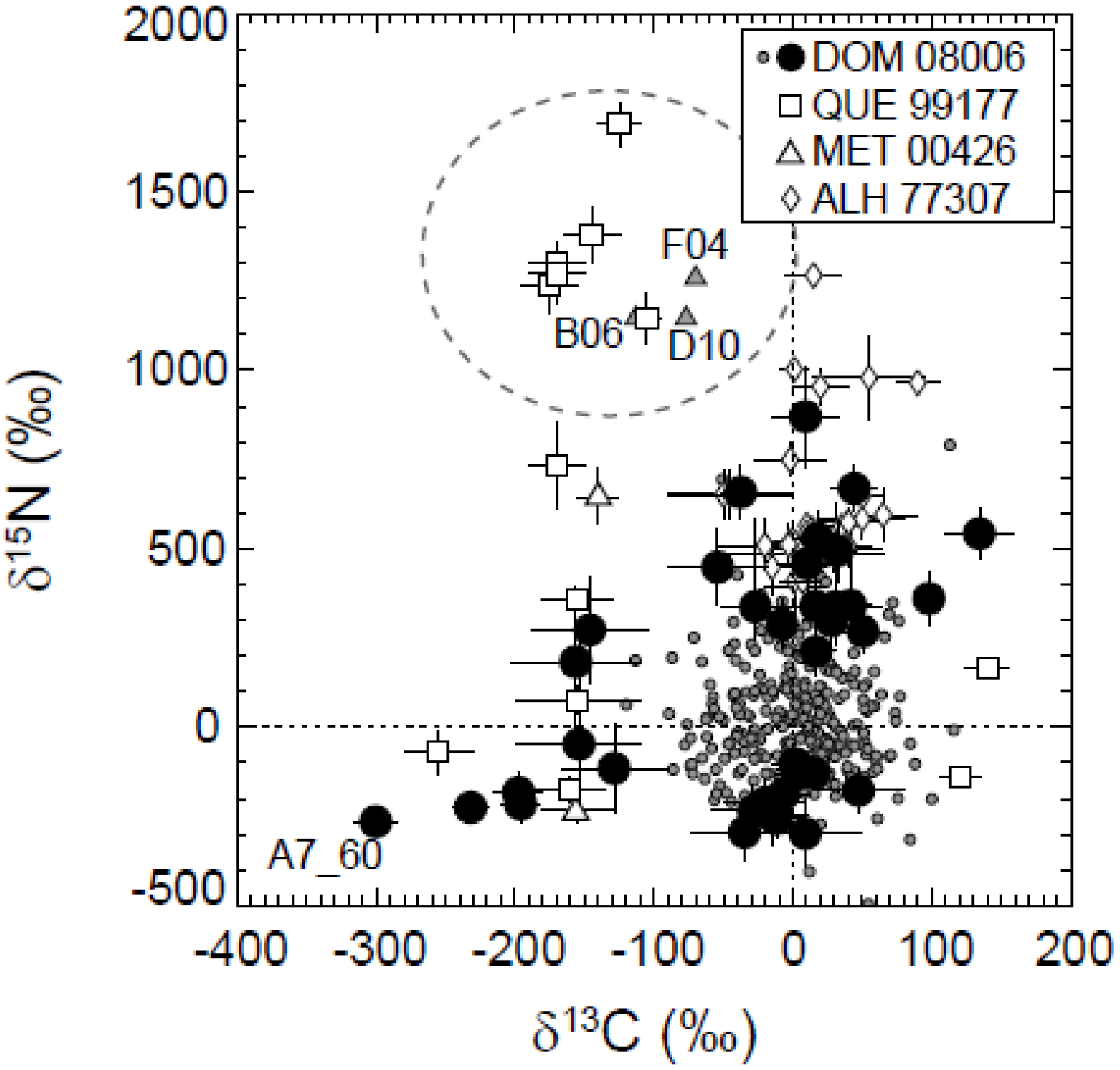

Figure 10. C- and N-isotopic ratios, expressed as δ-values, of organic grains that are anomalous in one or both of these elements from DOM 08006 (present study), QUE 99177 and MET 00426 (Floss and Stadermann, 2009b), and ALH 77307 (Bose et al., 2012). Small grey circles are DOM 08006 data that do not meet our criteria for being considered anomalous (see text). F04 indicates an anomalous grain found in an IDP (Floss et al., 2004), B06 indicates an isotopic hotspot in insoluble organic matter from a CR chondrite (Busemann et al., 2006), and D10 indicates a carbonaceous nanoglobule from comet Wild-2 (De Gregorio et al., 2010). The dotted lines indicate the terrestrial values. Dashed ellipse indicates population of $^{15}$N-enriched, $^{13}$C-depleted organic grains seen in several other extraterrestrial samples, but not DOM 08006. Isotopic images of grain A7_60 are shown in Figure 11.

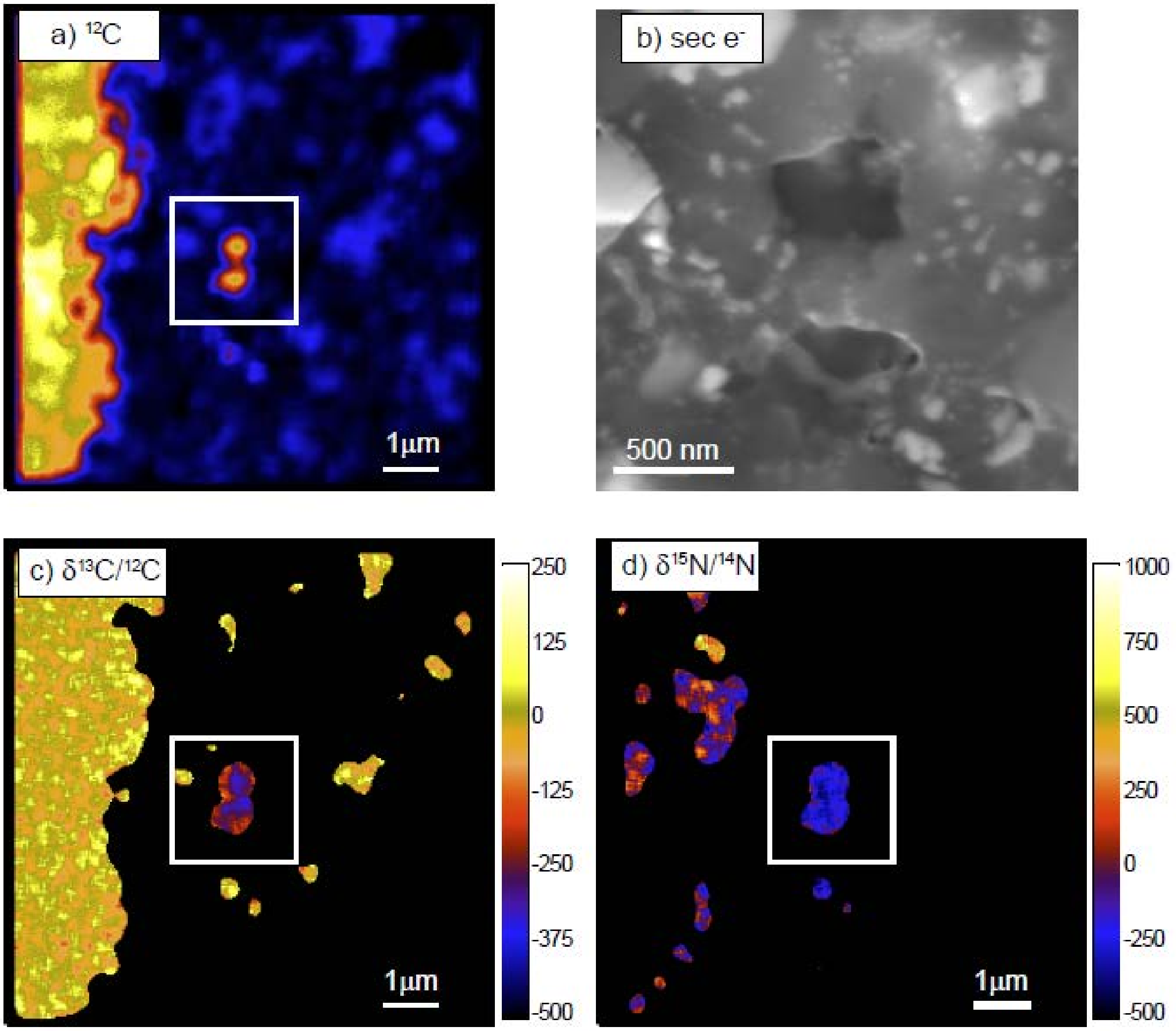

Figure 11. Images of DOM 08006 area (from Area 7, Figure 1) containing C- and N-anomalous grain A7_60 (white boxes). a) NanoSIMS $^{12}C$ image, b) secondary electron image, c) NanoSIMS $\delta^{13}C$ image, and d) NanoSIMS $\delta^{15}N$ image. Delta-values are in ‰.

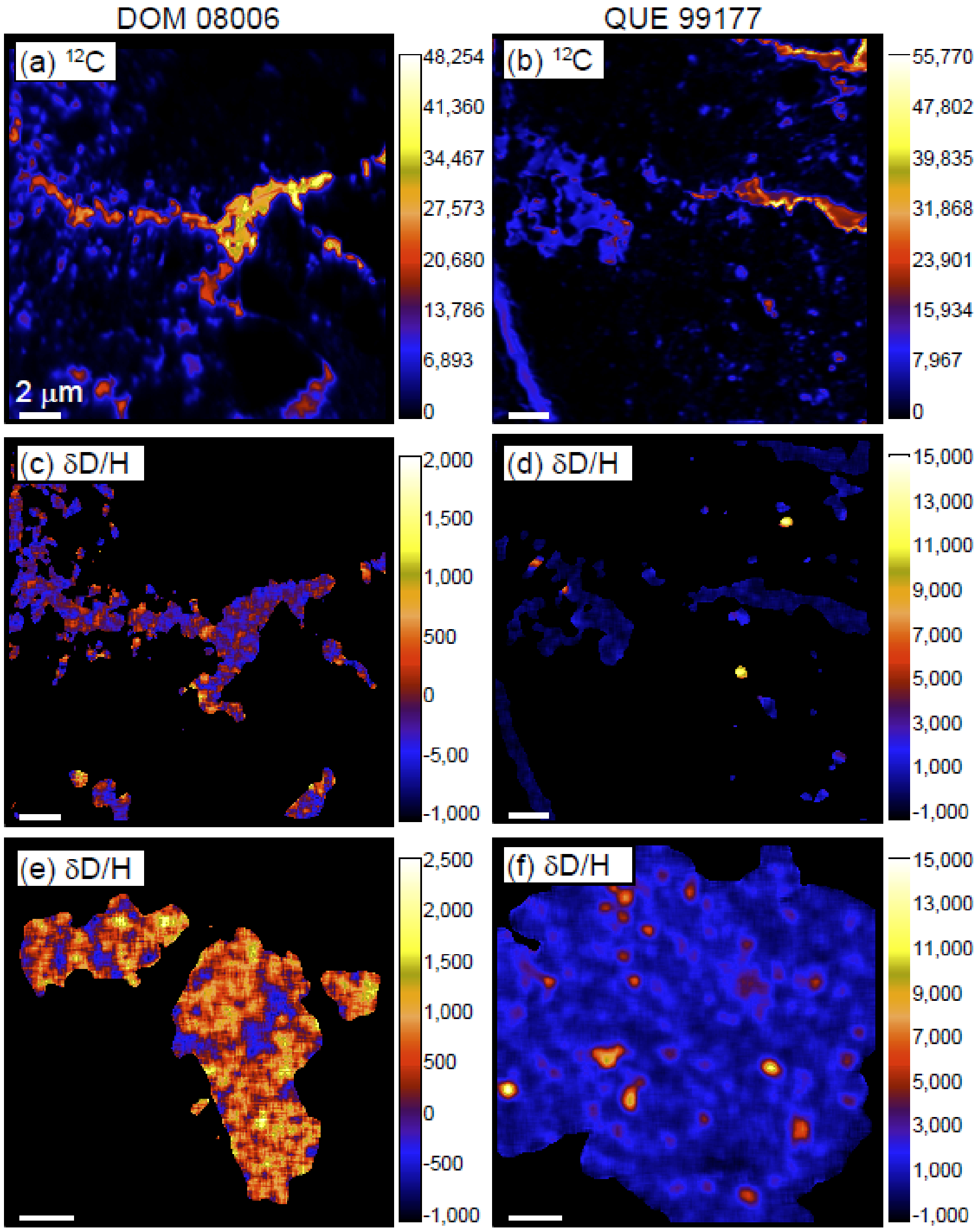

Figure 12. NanoSIMS images of DOM 08006 (left) and QUE 99177 (right). (a) and (b) $^{12}C$ ion images; units are counts per second. (c) and (d) In situ δD images; units in ‰. (e) and (f) Purified IOM δD images; units in ‰. Note the similar intensity scales for $^{12}C$ images but much wider scales for δD in the QUE 99177 data compared to the DOM 08006 data. The apparent isotopic variations in (c) and (e) are not statistically significant. QUE 99177 in situ data were acquired previously with the Carnegie NanoSIMS under similar conditions to those reported here for DOM 08006. Scale bars are 2 µm.

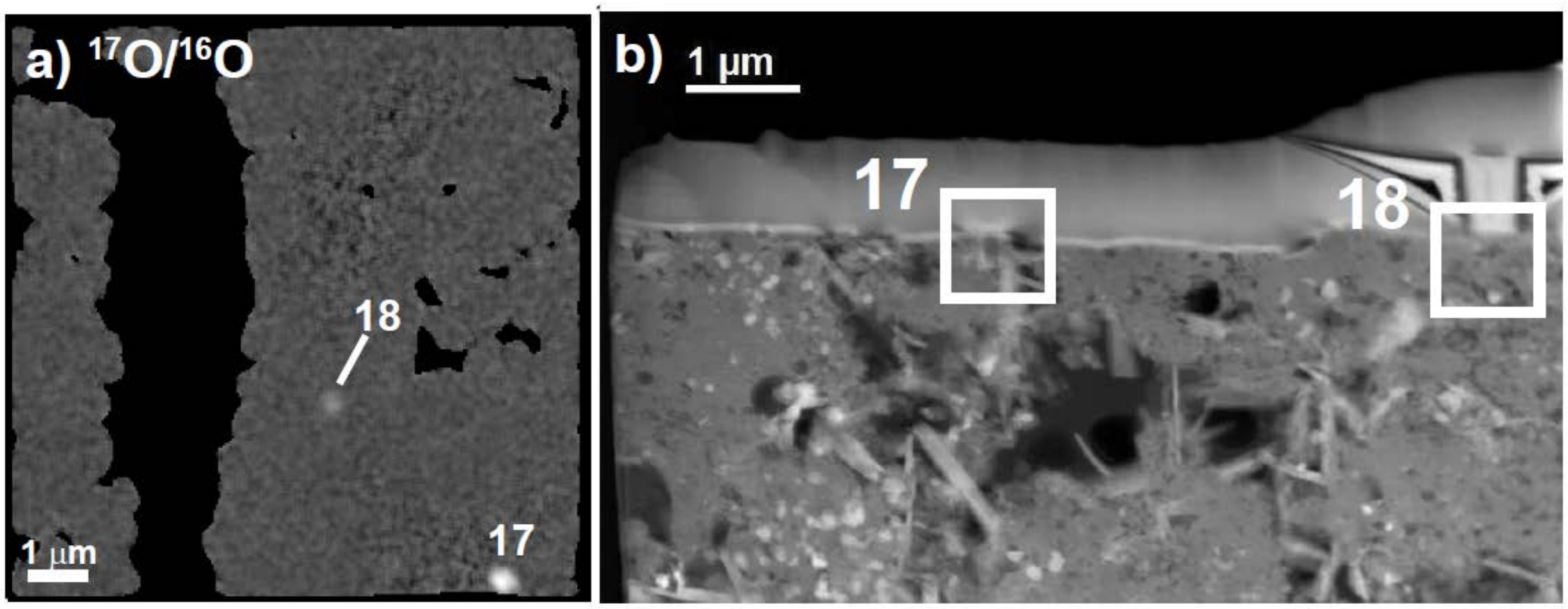

Figure 13. a) NanoSIMS $^{17}O/^{16}O$ map containing presolar grains DOM-17 and DOM-18. b) Bright-field TEM image of FIB section containing presolar grains.

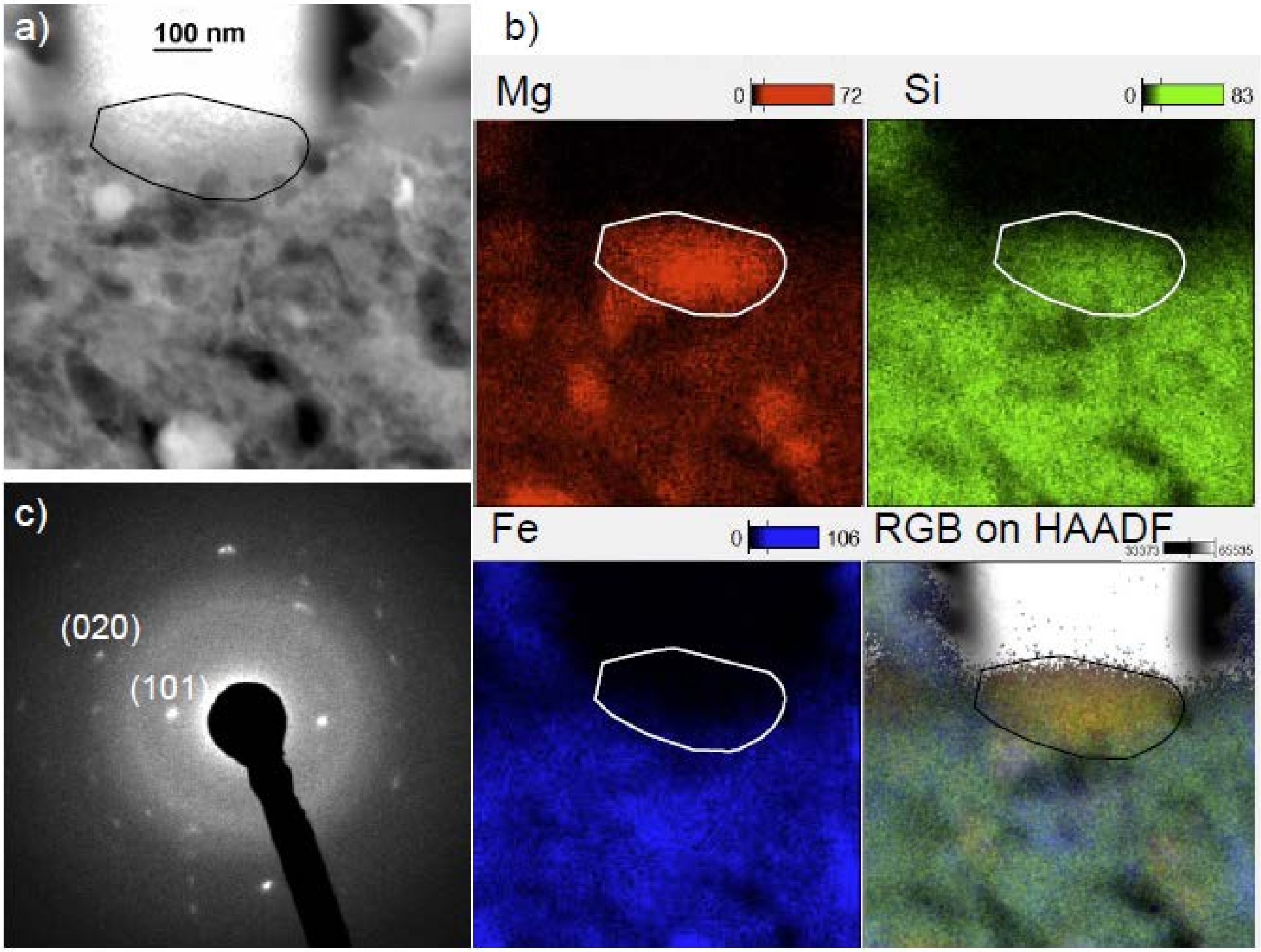

Figure 14. a) High Angle Annular Dark Field (HAADF) STEM image, b) STEM-based EDS element maps, and c) selected-area electron diffraction pattern of grain DOM-18. The outline of the grain is indicated on TEM image and EDS maps. The "RGB on HAADF" panel overlays a false-color combination of the Mg, Si, and Fe maps (red, green, blue, respectively) on the HAADF STEM image; the yellow color indicates grain is poor in Fe. The diffraction pattern indicates that it is crystalline olivine, with specific lattice planes indicated for two diffraction spots.

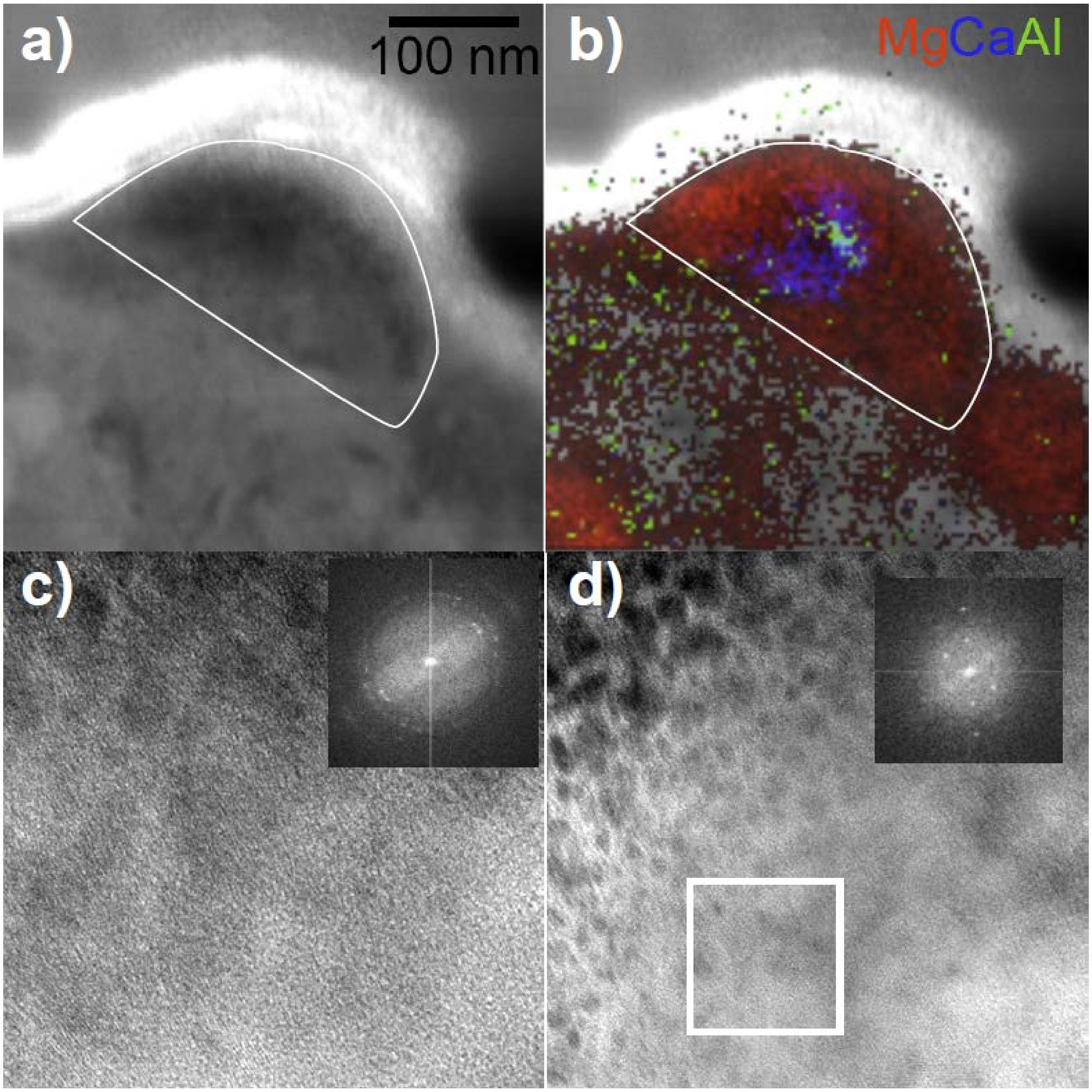

Figure 15. Transmission electron microscopy of presolar grain DOM-77: a) Dark field STEM image. b) RGB STEM-EDS overlay on the STEM image. c) HR-TEM and Fast Fourier Transform (FFT) of Mg-rich olivine region showing lattice spacings of 0.251 nm, 0.225 nm, and 0.205 nm. d) HR-TEM and Ca-rich region with lattice spacings 0.406 nm and 0.367 nm; FFT inset is from the area indicated by the box. Grain outline is indicated on a) and b).

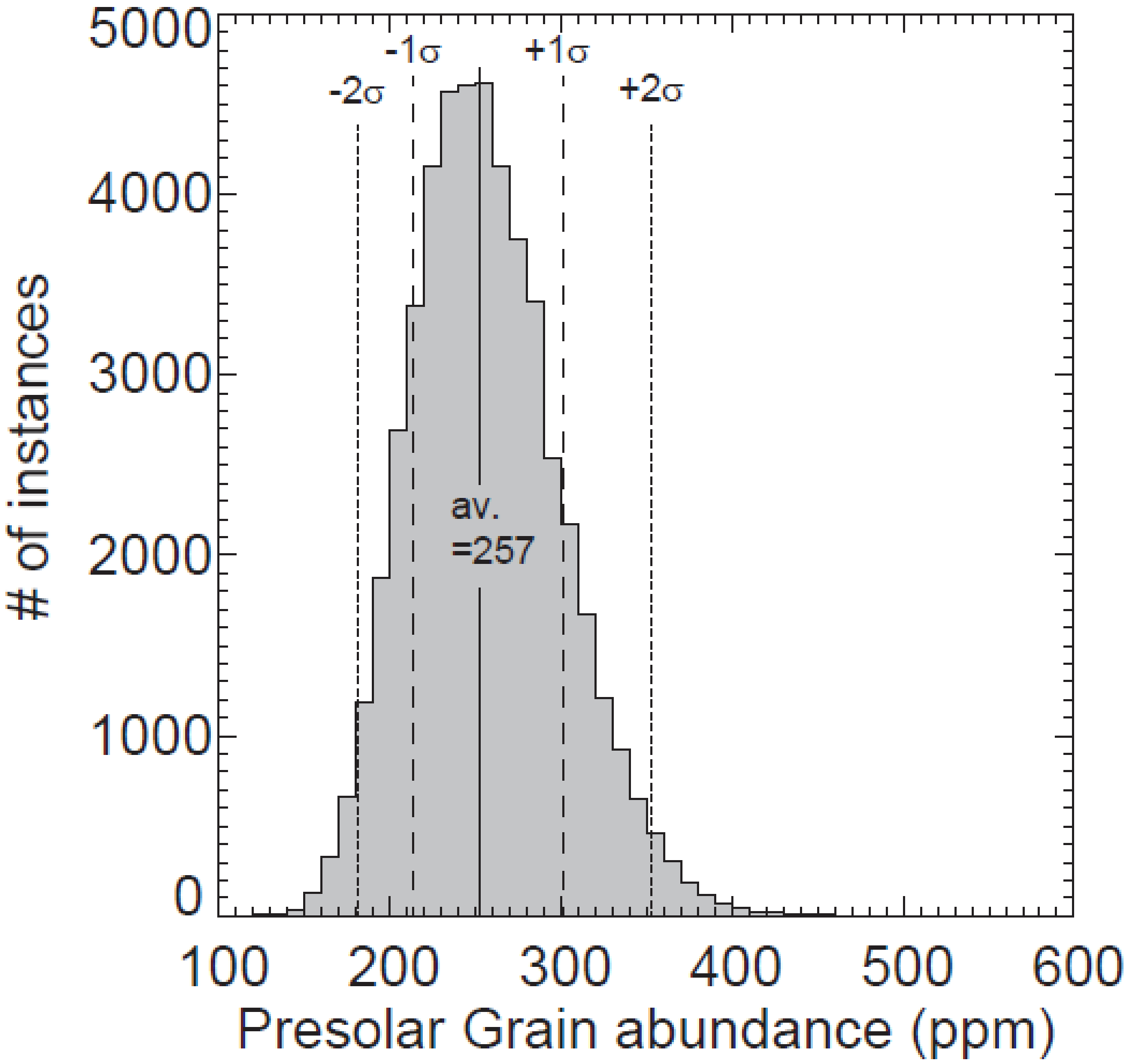

Figure 16. Results of a Monte Carlo calculation to estimate the uncertainty in the derived abundance of O-rich presolar grains in DOM 08006 (see text for details).

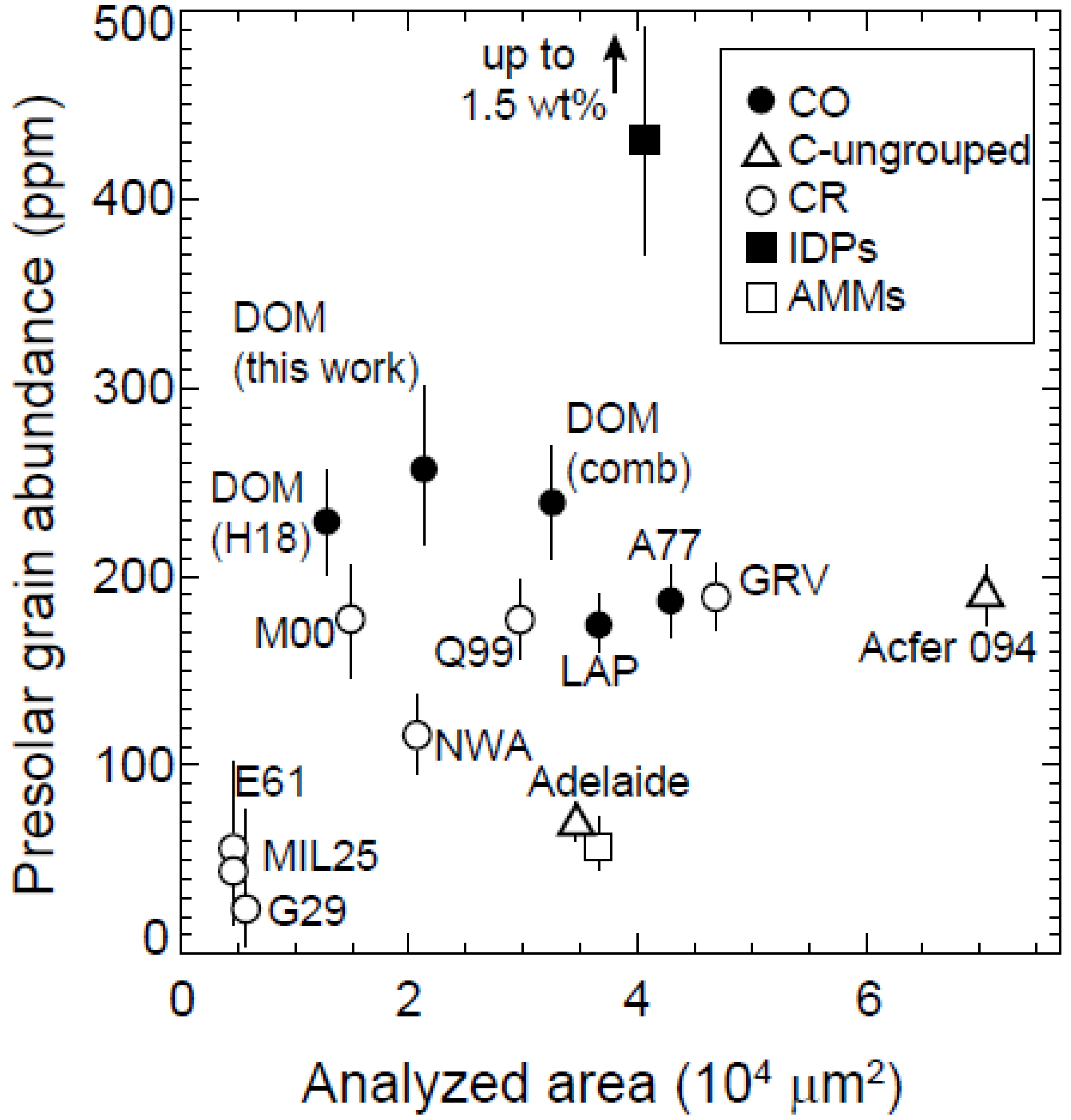

Figure 17. Abundances of presolar O-rich grains in primitive extraterrestrial materials plotted against total area mapped by NanoSIMS. DOM (this work): DOM 08006 data reported here; Acfer 094 (Vollmer et al., 2009b); Adelaide (Floss and Stadermann, 2012); A77: ALH 77307 (Nguyen et al., 2010); AMMs: Antarctic Micrometeorites (Yada et al., 2008); DOM (H18): DOM 08006 matrix (Haenecour et al., 2018); E61: Elephant Moraine 92161 (Leitner et al., 2016); DOM (comb): combined DOM 08006 dataset from the present study and that of Haenecour et al. (2018); GRV: Grove Mountains 02170 (Zhao et al., 2013); G29: Graves Nunataks 95229 (Leitner et al., 2016); IDPs: interplanetary dust particles (Floss et al., 2006; Busemann et al., 2009); LAP: LAP 031117 matrix (Haenecour et al., 2018); MIL25: Miller Range 07525 (Leitner et al., 2016); M00: MET 00426 (Floss and Stadermann, 2009a); NWA: Northwest Africa 852 (Leitner et al., 2012b); Q99: QUE 99177 (Floss and Stadermann, 2009a; Nguyen et al., 2010). Error bars are 1σ.